\documentclass[prd,superscriptaddress,twocolumn,floatfix,preprintnumbers,altaffilletter, letterpaper]{revtex4}

\usepackage{amsmath,amssymb,amsfonts}   % for math

\usepackage{xcolor}
\usepackage{booktabs}
\usepackage[colorlinks=true]{hyperref}
\hypersetup{citecolor=teal}
\usepackage[T1]{fontenc}

\usepackage{url}

\usepackage{graphicx}  % needed for figures

\begin{document}

 \preprint{IUCAA-06/2018}
 \preprint{LIGO-P1800289}

\title{Effect of Induced Seismicity on Advanced Gravitational Wave Interferometers}

\author{N Mukund} 
\email{nikhil@iucaa.in}
\affiliation{Inter-University Centre for Astronomy and Astrophysics (IUCAA), Post Bag 4, Ganeshkhind, Pune 411 007, India}
\author{B O'Reilly}\email{brian@ligo-la.caltech.edu}
\affiliation{LIGO Livingston Observatory, Livingston, LA 70754, USA}
\author{S Somala}\email{surendra@iith.ac.in}
\affiliation{Department of Civil Engineering, Indian Institute of Technology (IIT) Hyderabad, Hyderabad, India 502285}

\author{S Mitra}\email{sanjit@iucaa.in}
\affiliation{Inter-University Centre for Astronomy and Astrophysics (IUCAA), Post Bag 4, Ganeshkhind, Pune 411 007, India}

\date{\today}
\date{\today}

\begin{abstract}
Advanced LIGO and the next generation of ground-based detectors aim to capture many more binary coalescences through improving sensitivity and duty cycle. Earthquakes have always been a limiting factor at low frequency where neither the pendulum suspension nor the active controls provide sufficient isolation to the test mass mirrors. Several control strategies have been proposed to reduce the impact of teleseismic events by switching to a robust configuration with less aggressive feedback. The continental United States has witnessed a huge increase in the number of induced earthquake events primarily associated with hydraulic fracking-related waste water re-injection. Effects from these differ from teleseismic earthquakes primarily because of their depth which is in turn linked to their triggering mechanism. In this paper, we discuss the impact caused due to these low magnitude regional earthquakes and explore ways to minimize the impact of induced seismicity on the detector.
\end{abstract}

\maketitle

\section{\label{sec:ChD:Intro}Introduction}

Terrestrial gravitational wave (GW) observatories like Advanced LIGO and Advanced Virgo have detected coalescing compact binaries consisting of black holes and neutron stars~\citep{AbEA2016a,AbEA2016e,AbEA2017a,AbEA2017b,AbEA2017c} and efforts are in progress to achieve the design sensitivity of the detectors. This process requires correctly understanding the limiting noise sources and developing strategies to mitigate them. At frequencies below 30 Hz, ground vibrations significantly affect the duty cycle of the LIGO instruments and the quality of the strain data used for GW analysis.  The passive and active isolation systems installed at Advanced LIGO are designed to provide seismic noise suppression at frequencies above 1 Hz \citep{MaLa2015}. Seismic isolation from the micro-seism is achieved using feedforward subtraction, but there is typically no isolation (and in some cases amplification), below 0.1 Hz. 
Earthquakes contribute the most to this frequency band and previous studies \citep{BiWa2018} associate a high probability for the interferometer to go to an uncontrolled state (lockloss) whenever the resulting ground motion rises above few microns per second. This instability is partly due to the self-inflicted gain peaking of the seismic isolation stage which at these low frequencies leads to platform motion which is quite often higher than the actual ground motion. Earthquakes thus significantly contributor to detector downtime and the resulting recovery period can vary from a few minutes to hours. SEISMON \citep{seismon1,seismon2} the low latency early earthquake warning (EEW) application installed at the GW detectors focuses on high magnitude teleseismic events. The alert issued based on information received from the USGS seismological network has a latency of approximately ten minutes and includes information about the time of arrival and the expected ground motion from the Rayleigh waves. Such alerts allow preparation of the instrument to stay operational during the quake or to put it in a safe condition. For the first time, we discuss the impact of induced low magnitude earthquakes($M_{w}$ < 5) of regional origin on current and next-generation GW interferometers. 

The paper is organized as follows, Section~\ref{sec:INDUCED} discusses induced seismicity and the sudden increase in regional earthquakes in the conterminous US. Section~\ref{sec:OKH} focuses specifically on the Oklahoma region and how events from this region drive the interferometer to a state of lockloss. In Section~\ref{sec:MODEL}, we describe the results from simulated events from the Tuscaloosa Marine Shale (TMS) region and look into the proximity effects. Finally in Section~\ref{sec:IMPACT}, we briefly discuss how this new source of noise could be detrimental to the future generation of terrestrial detectors and possible strategies to minimize the impact. 

\begin{figure*}[!htb]
\begin{center}
\includegraphics[width=0.75\linewidth]{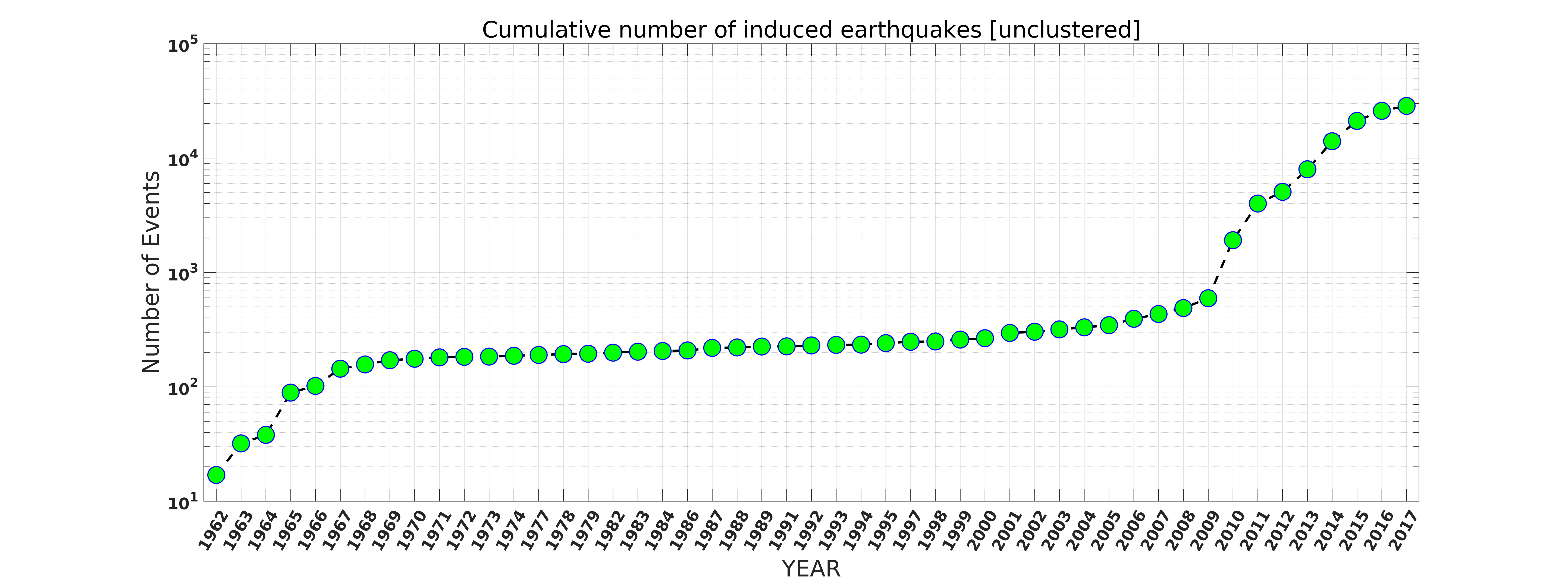}
\caption{\label{fig:Cumulative_EQs}Cumulative distribution of induced earthquakes in conterminous US ( Data obtained from U.S. Geological Survey data release,2018 \citep{induced_seismicity_catalog}). The drastic increase since 2008 strongly correlates with the increased activities related to hydraulic fracking.}
\end{center}
\end{figure*}

\begin{figure*}[!htb]
\centering
\includegraphics[width=0.75\textwidth]{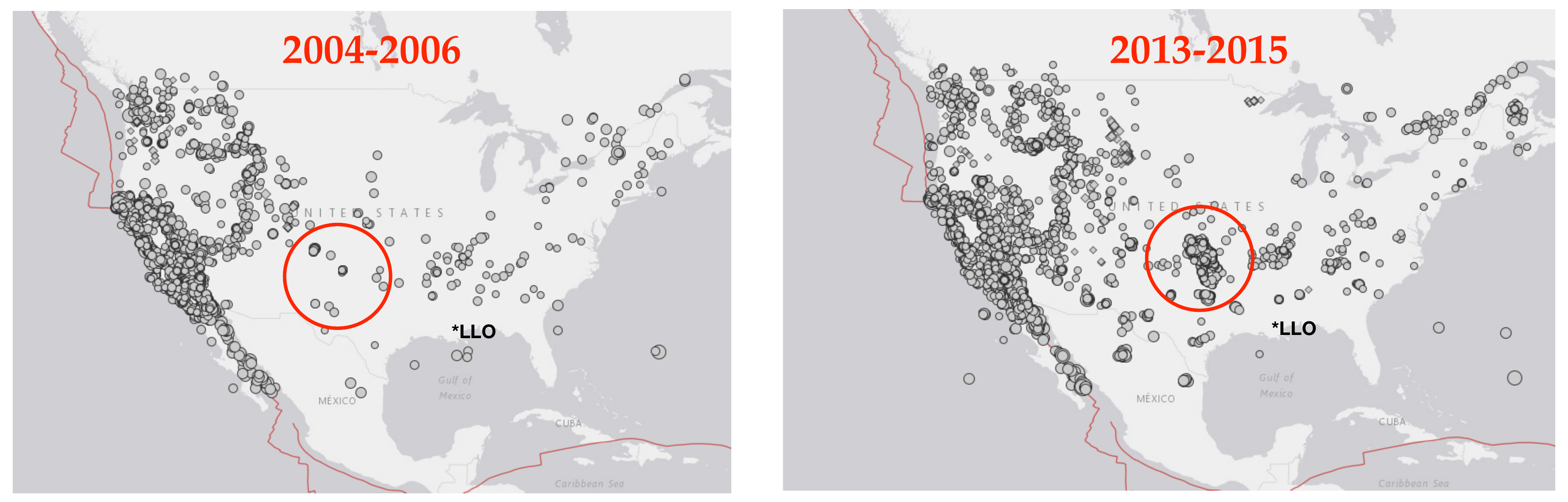}
\caption{\label{fig:OklahomaEvents} Plot comparing the earthquakes in Oklahoma during 2004-06 and 2013-15 (data obtained from USGS website).}
\end{figure*}

\section{Induced Seismicity}\label{sec:INDUCED}

Induced seismicity refers to low magnitude earthquakes caused by industrial activity that stresses the fault structures resulting in a rupture. Earthquakes happen when there is a slip in the fault resulting in a release of stored stress energy. Rupture is initiated when the effective shear stress exceeds the frictional resistance, and the energy gets radiated away as seismic waves. Shallow earthquakes typically occur from a slip along an existing fault line. Fault rupture occurs when the shear stress between the foot-wall and the hanging wall exceeds the effective normal stress given by the Mohr-Coulomb failure criterion\citep{scholz2002mechanics,zoback2010reservoir,national2013induced}, 
\begin{equation}
\tau_{c} = C + \mu (\sigma_{n} - P) \,,
\end{equation}
where $\tau_{c}$ is the critical stress, $C$ is the cohesion, $\mu$ is the coefficient of friction, $\sigma_{n}$ is the normal stress, $P$ is the pore pressure due to the fluid infiltration between the fault \& neighboring rock. The magnitude of the resulting earthquake depends on the stress levels and the size of the fault over which they are applied. Fault rupture leads to body waves composed of primary pressure waves or P-waves and secondary shear waves or S-waves. Based on the particle motion, either along the horizontal or the vertical, S-waves can be further classified into SH or SV waves. Surface wave generation happens from the complex interaction of S-waves with the Earth's free surface. Constructive interference of multiple SH modes results in surface Love waves \citep{aki2002quantitative} which are prominently captured in the horizontal component of seismometer data. Rayleigh waves, observed in the vertical seismic spectrum, are created due to the interaction of SV waves with the P-waves at the free surface and causes elliptical motion in the vertical plane with a vertical to horizontal amplitude ratio close to 3:2.

The last several years have seen an unprecedented increase in the number of induced earthquakes in the conterminous US \citep{hough2014shaking,ellsworth2015increasing} from regions that were previously seismically quiet (See Fig. \ref{fig:Cumulative_EQs}). The Oklahoma region is the most striking example where the rise in low magnitude earthquakes has been pronounced since 2008 \citep{induced_seismicity_catalog}. This phenomenon is strongly correlated with the rise in fracking activities within the state as part of the growing oil industry. Hydraulic fracking is the process of horizontal drilling followed by injection of liquid at high pressure. This process results in excess of wastewater which is hazardous in general and so needs to be disposed of through re-injection back to the crust. Injection fluid can directly get in contact with the fault structure resulting in an increase in pore pressure, or it could lead to a change in the loading conditions on the fault leading to a tremor. The proximity of the epicenter to the interferometer site implies even minor quakes from these regions could be a potential cause of lockloss at the interferometers. This coupled with a high event rate could hamper the day-to-day operations of the detector.  Factors that determine induced seismicity are still an active area of research \citep{weingarten2015high}, but, in general, they include the number of injection wells, the rate of injection, wellhead injection pressure, proximity to the crystalline basin and cumulative injection volume. The triggered low magnitude earthquakes usually do not result in structural damage as the induced motion is of orders of tens of micrometers but is sufficient to disrupt the functioning of a GW interferometer. 

\begin{figure*}[!htb]
\centering
\includegraphics[width=0.8\textwidth]{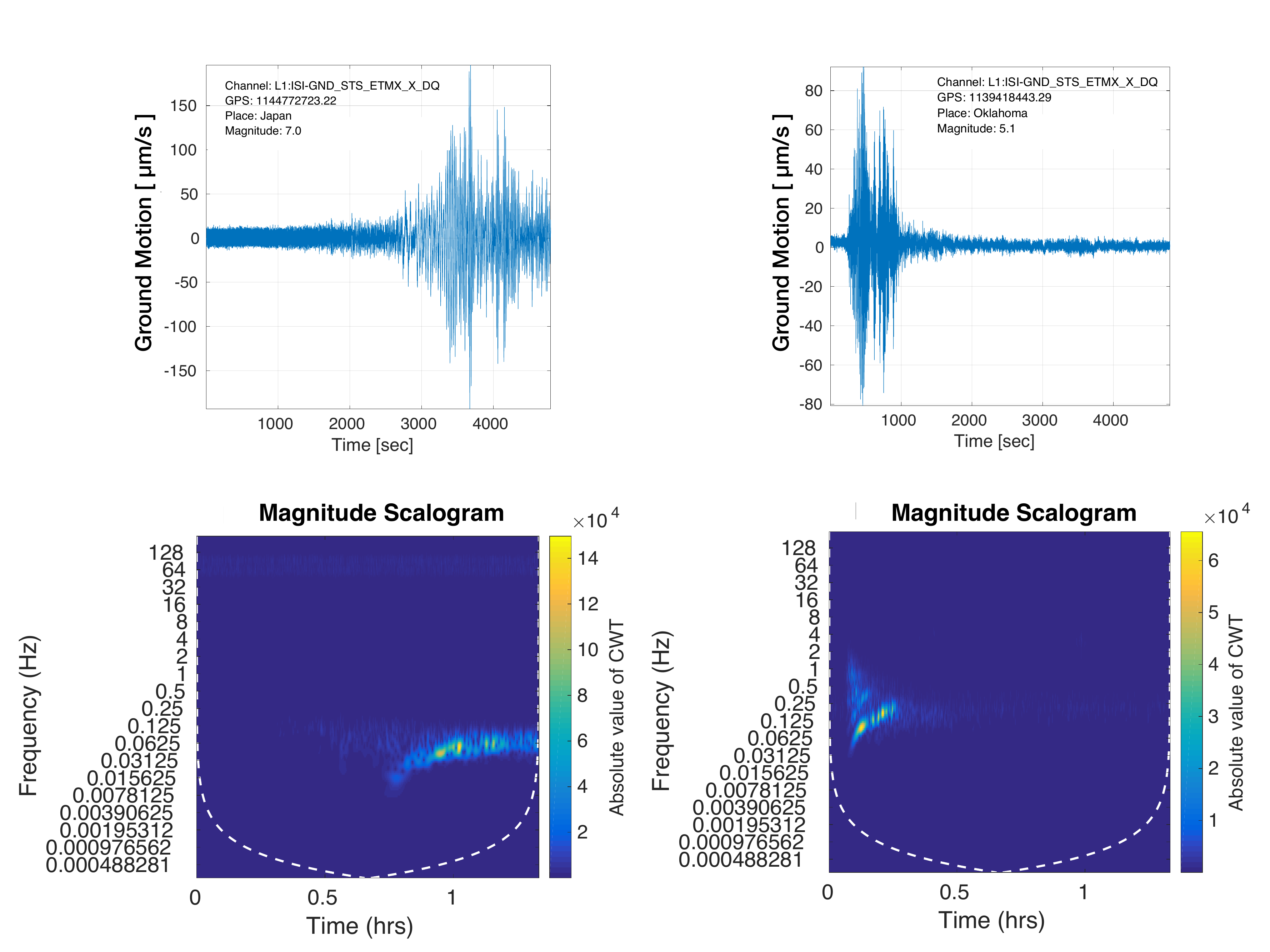}
\caption{Continuous Wavelet Transform (CWT) based time-frequency map of ground motion recorded along the horizontal east-west direction observed at the Livingston Observatory. The left plot shows a 4.3 magnitude induced earthquake from Oklahoma (2014-07-29 02:46:36 UTC) while the right plot is for a seven magnitude earthquake from Japan (2016-04-16 16:25:00 UTC). Shallowness in depth and nearness to site causes the Oklahoma event to produce significant motion at higher frequencies.}
\label{fig:EQ_comparison} 
\end{figure*}

\section{Oklahoma Events}\label{sec:OKH}
\begin{figure*}[!htb]
\centering
\includegraphics[width=0.8\textwidth]{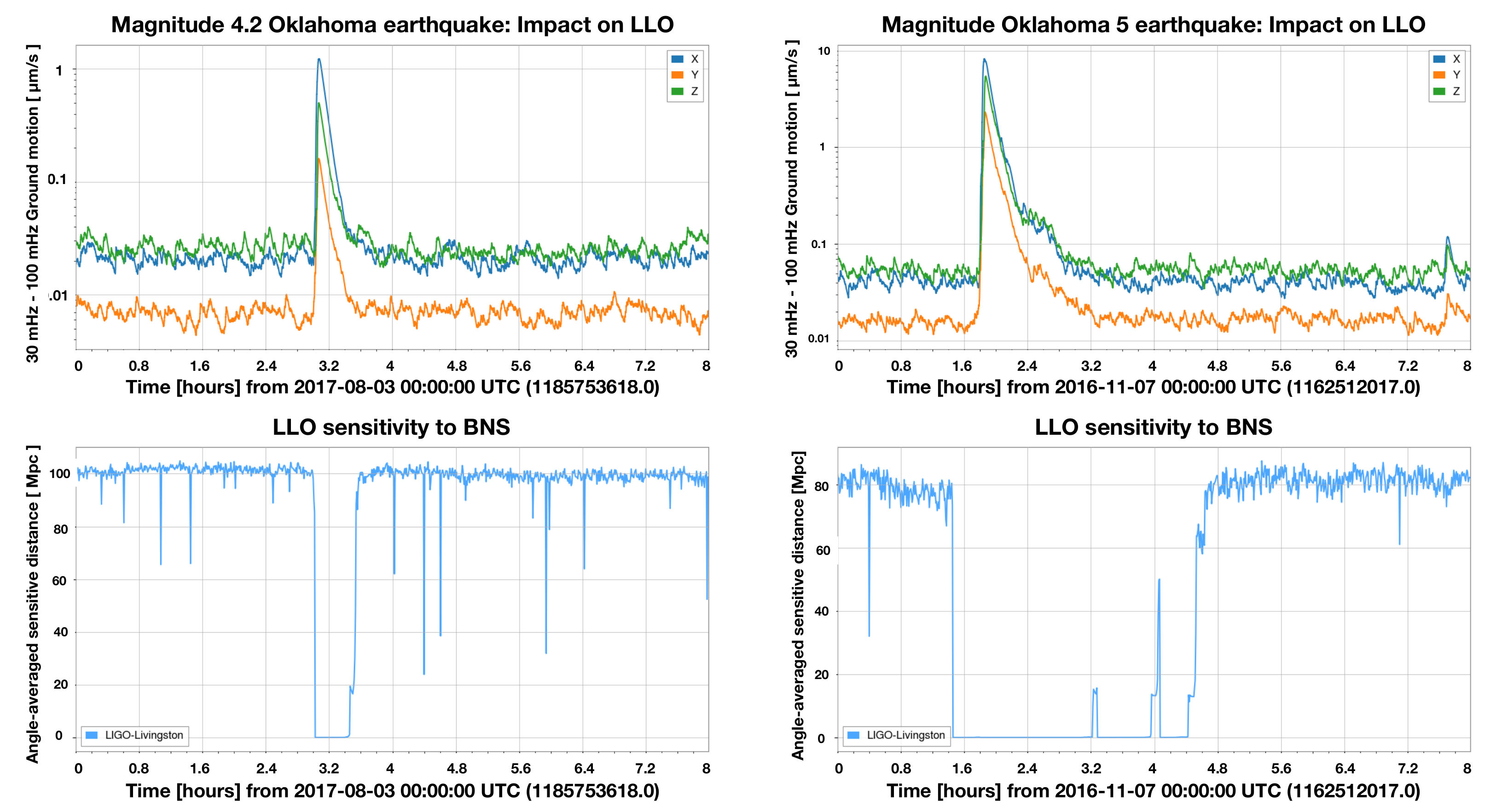}
\caption{ Effect of induced earthquakes on the LIGO Livingston detector. The top row shows the band-limited RMS motion recorded by seismometers kept at X-end, Y-end \& the center station. The bottom row provides the average sensitivity in terms of distance (averaged across all-sky positions and orbital inclinations) to which LLO can detect a signal from the merger of 1.4 - 1.4 solar mass binary neutron star with an SNR of 8. The first column depicts a 4.2 magnitude earthquake from the state of Oklahoma which lead to a downtime of thirty minutes. The effect is more severe with the second one of magnitude five from the same state where the instrument is seen to take more than three hours to be back in operation. Two unsuccessful re-locking attempts carried out during the downtime can be clearly from the last subplot.}
\label{fig:Oklahoma_lockloss}
\end{figure*}
\begin{figure*}[!htb]
\centering
\includegraphics[width=0.85\textwidth]{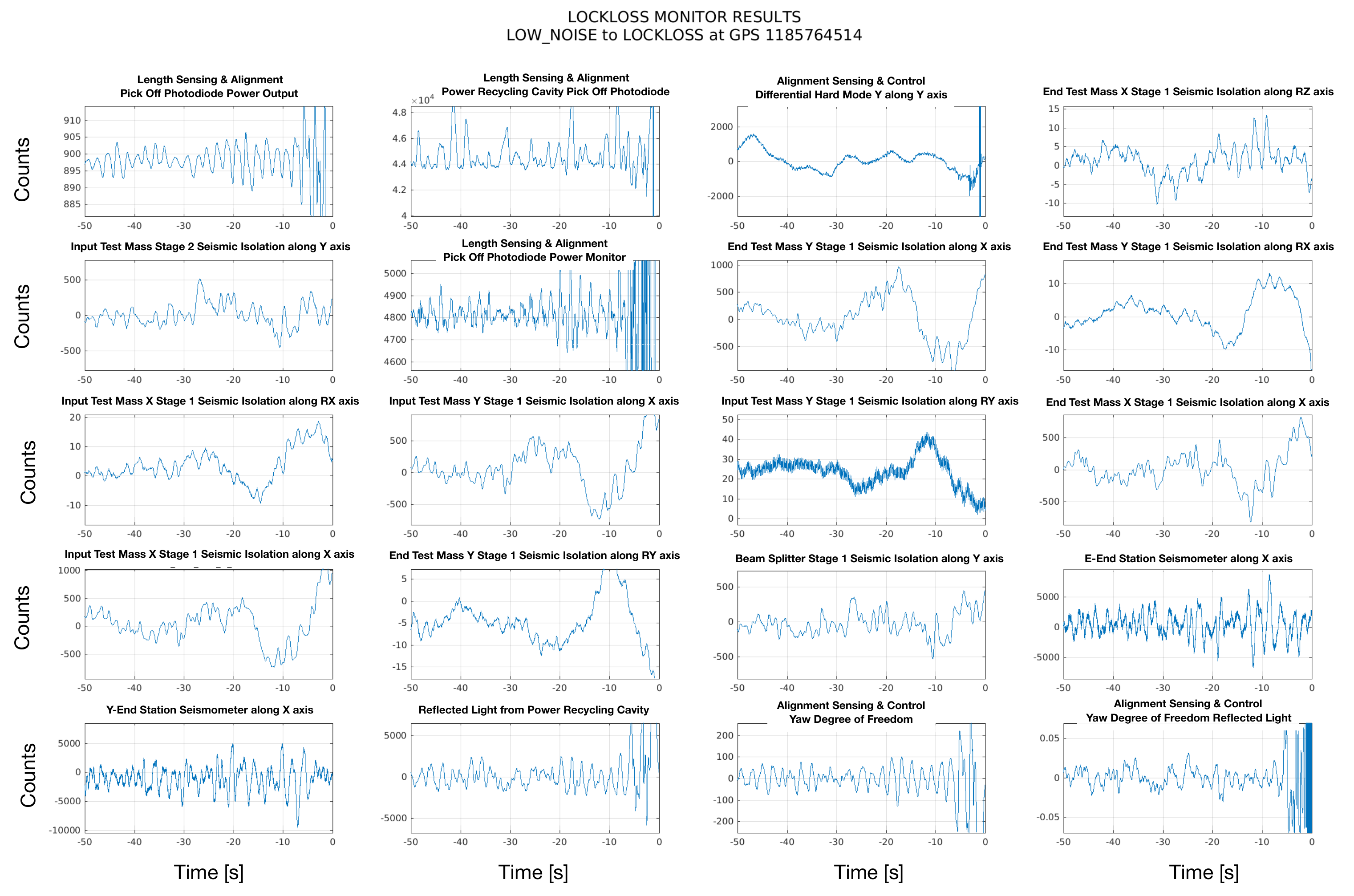}
\caption{Effect of 4.2 magnitude Oklahoma earthquake on various LIGO subsystems. Growing oscillations are visible in all the corner and end station vibration isolation control loops tens of seconds prior to lockloss. }
\label{fig:lockloss}
\end{figure*}

\begin{figure*}[!htb]
\centering
\includegraphics[width=0.75\textwidth]{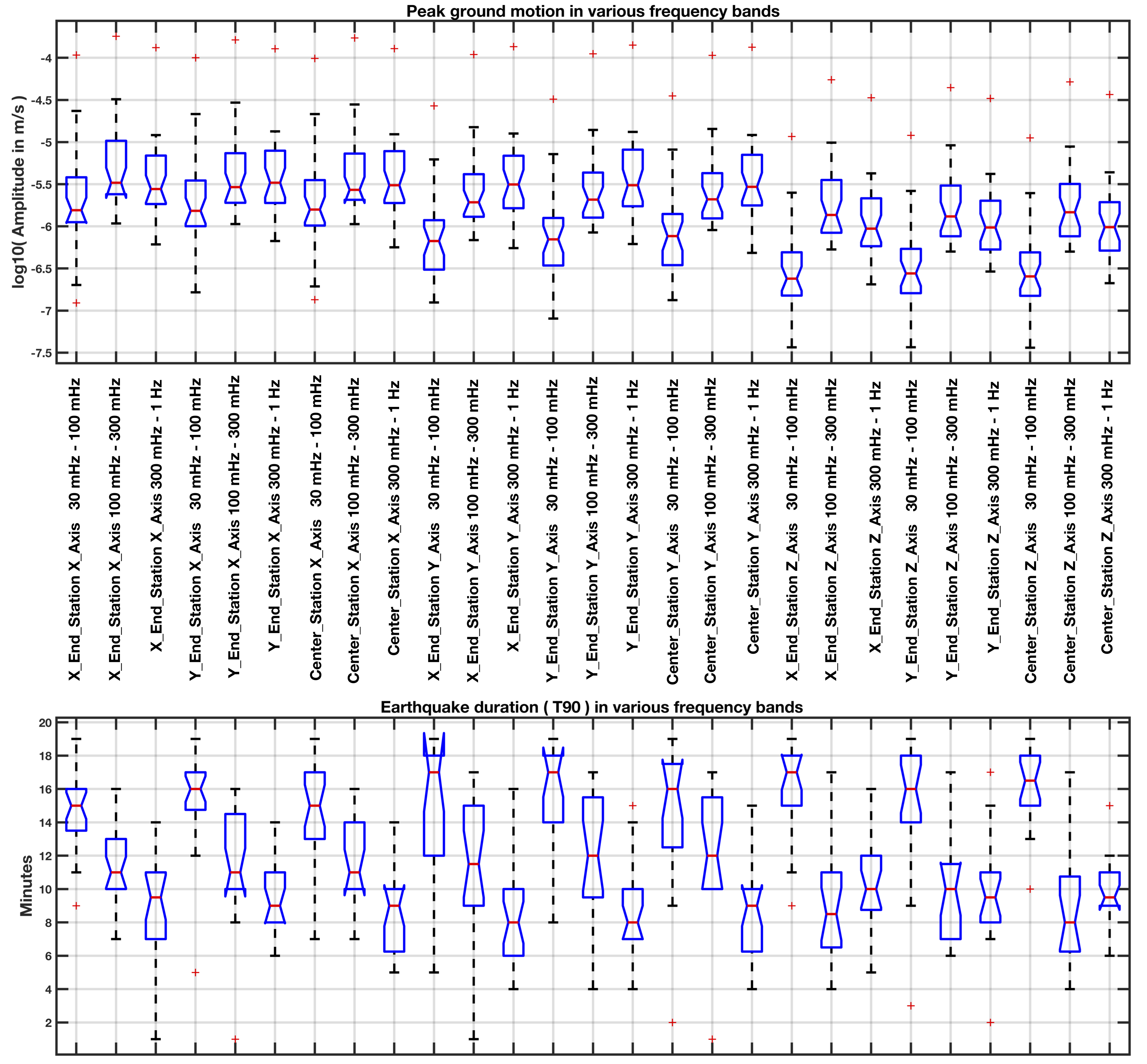}
\caption{Box plots showing the distribution of peak ground motion and duration of 31 Oklahoma earthquakes from 2015 to 2018 with magnitude above 4.2 (see Appendix \ref{App2} for individual event information) as seen in multiple seismometer at different frequency bands. }
\label{fig:BoxPlots} 
\end{figure*}
Oklahoma has been experiencing a series of earthquakes of magnitude above 4 in the past few years (See Fig.~\ref{fig:OklahomaEvents}), and several previous studies have associated these with wastewater injection activities in the state \citep{keranen2014sharp,mcnamara2015earthquake,mcnamara2015reactivated}. 

In Fig.~\ref{fig:EQ_comparison}, we compare a 5.1 magnitude induced earthquake from Oklahoma with a magnitude seven teleseismic event from Japan. The time-frequency plot clearly shows the difference between the two both in frequency content, duration and time of arrival.  A magnitude 5.0 earthquake that occurred at Oklahoma on 7th November 2016 01:44:24 UTC lead to the first-ever reported lockloss at LIGO Livingston Observatory (LLO) from induced earthquakes (See Fig.~\ref{fig:Oklahoma_lockloss}). The latency of the early warning system currently installed at the site is around 10 minutes which is greater than the arrival time of Rayleigh waves from regional sites like Oklahoma. The event led to a Rayleigh wave amplitude of 11 microns/sec and saturated most of the suspension models and possibly the seismometers. The lockloss happened within 5 minutes of the event which was less than the time it took to show up as an alert in USGS and subsequently in SEISMON. There was a delay of 3 hrs before the interferometer could be successfully locked again although it usually takes around 30-45 minutes to recover from such a quake. Our studies show that any earthquake from Oklahoma with a magnitude above 4.2 and occurring at a depth around 5 km has the potential to cause lockloss at the site. We currently have very limited information on how the lockloss happens, where the exact failure point is, or if it is the same for all the earthquakes. Advanced LIGO which is a Dual-Recycled Fabry-Perot Michelson Interferometer makes use of an elaborate hierarchical control strategy to achieve and maintain cavity resonance \citep{0264-9381-31-24-245010}. This process involves bringing multiple degrees of freedom to a linear operating point, and usually the associated error signals have a very narrow linear regime. The excess ground motion usually triggers instabilities in the control signals which either tend to drive the digital to analog converters beyond their actuation limit or causes short duration transients or glitches finally leading to loss of resonance in the optical cavities. In Fig.~\ref{fig:lockloss} we show the behavior of various subsystems (including Alignment Sensing $\&$ Control [\textbf{ASC}], Internal Seismic Isolation [\textbf{ISI}] and Length Sensing $\&$ Control [\textbf{LSC}] ) seconds prior to a typical earthquake induced lockloss.

Table~\ref{table:Oklahoma_events} given in Appendix~\ref{App2} lists 31 earthquakes with magnitude greater than 4.2 that occurred in the Oklahoma region from 2013 to 2018. 
In Fig.~\ref{fig:BoxPlots} we show the distribution of peak ground motion and duration of these events using data from three STS broadband seismometers (ETMX, ETMY, ITMY) located near the corner and end stations of LLO. Data is analyzed in various frequency bands, $30-100$~mHz (band I), $100-300$~mHz (band II) and $300-1000$~mHz (band III) to get a better understanding of these shallow regional earthquakes. Within the LIGO community, bands {I,II,III} are respectively referred to as earthquake, microseismic and ground motion bands.  We estimate duration in terms of T90 which provides the time interval for which the cumulative value of the band-limited spectral amplitude increases from 5\% to 95\% of its total value.

The earthquake band is observed to witness a median ground velocity of order $1 \;\mu m/s$ with the highest amplitude along the X-arm direction, followed by Y-arm and Z (vertical) direction. This suggests a dominant contribution from surface waves with significant contribution from the Love waves. The significant motion along the horizontal axis as seen from the sensors is undesirable as it could lead to misalignment of the optics, potentially increasing the time to lock the cavity again after an earthquake-induced lockloss. Interestingly, the micro-seismic and ground motion bands show even higher levels of motion, making these events different from the typical teleseismic events observed at LIGO \citep{seismon1}.  We also see a motion of micro radian for the pitch and yaw degrees of freedom associated with the pendulum suspensions. If the test mass mirrors swing with high amplitude, they become susceptible to stray charge deposition from contact with the suspension cage or the earthquake stoppers. This excess charge consequently leads to noise in the electrostatic actuation process which is used to control the various degrees of freedom of these mirrors. The increased seismic motion could also couple in a nonlinear fashion to the GW strain channel and amplify scattering noise in $50-200$~Hz. Such unwanted effects make it difficult to attain the desired sensitivity required for gravitational wave detection.

\section{Modeling Expected Ground Motion}\label{sec:MODEL}
\label{sec:II}
\begin{figure}[!htb]
\centering
\includegraphics[width=0.35\textwidth]{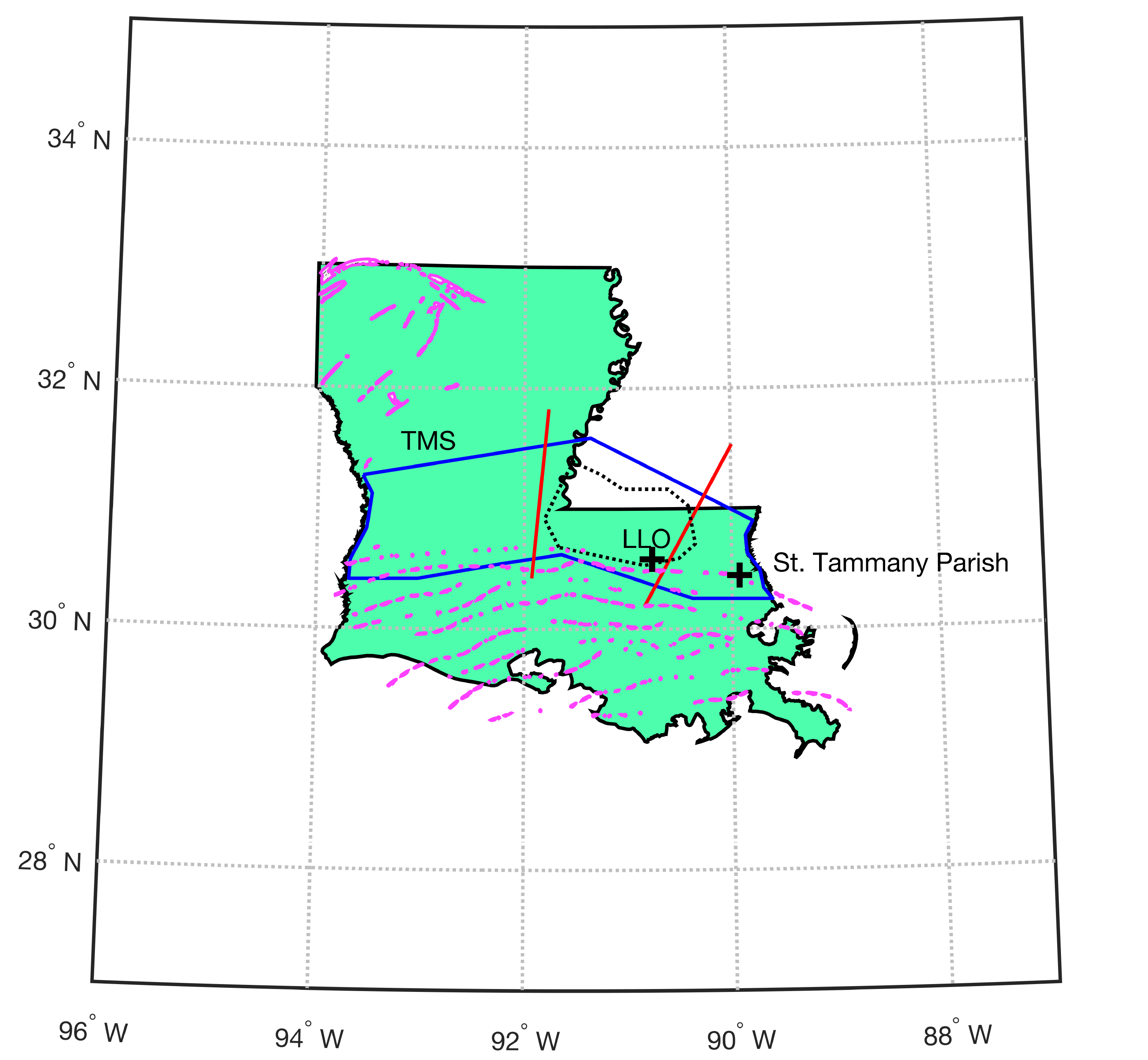}
\caption{Map of Louisiana overlaid with known active faults. Tuscaloosa Marine Shale region enclosed within the blue polygon. The region within the TMS having high resistivity is indicated by the dotted line. High resistivity is indicative of the presence of hydrocarbons making the region ideal for fracking.}
\label{fig:TMS} 
\end{figure}

Robust modeling is necessary to foresee the effects of possible future events, insights from which can be incorporated in future site selection studies. In this section, we estimate the expected ground motion caused due to induced seismicity in two different ways: first through analytical modeling based on past earthquake data, and secondly using a forward simulation of fault rupture obtained via spectral finite element modeling. We use the region around St. Tammany Parish, Louisiana, approximately 60 miles west of LLO. It is part of the Tuscaloosa Marine Shale (TMS) region with oil deposits trapped at 11,000-15,000 feet. Fig.~\ref{fig:TMS} shows the TMS region along with the active fault structure in the state of Louisiana \citep{Louisiana_Active_Faults}. As technology and techniques for fracking evolve, it may become economically viable in the near future to frack this shale formation \citep{Rigzone,NaturalGasWorld,DNR_Louisiana}. Louisiana already has many wastewater disposal wells, but the concern is that fracking disposal volumes are much higher and there is still a lot of uncertainty in the mechanism of inducing earthquakes. Previously, the Haynesville shale in Northern Louisiana has been fracked for several years, but since the activity peaked during the installation of advanced LIGO, we do not have sufficient evidence to access its impact.

\begin{figure*}[!htb]
\centering
\includegraphics[width=0.95\textwidth]{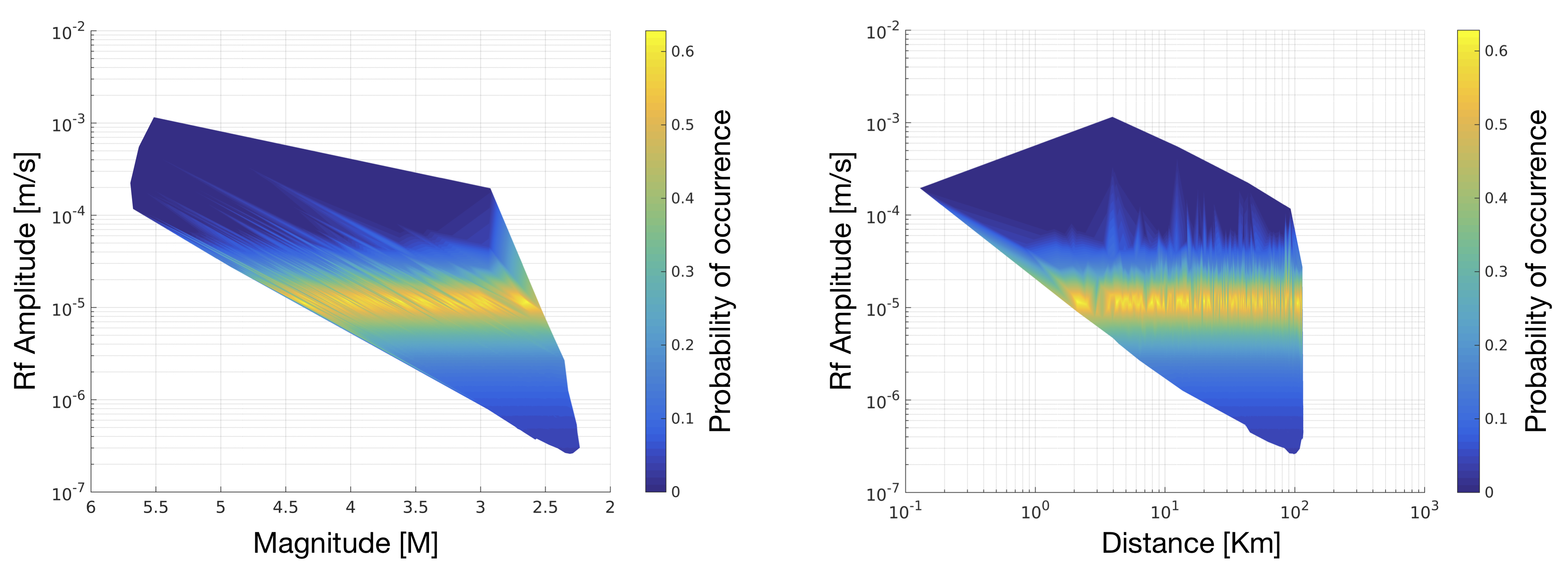}
\caption{Simulation of expected ground motion at LLO from induced Earthquakes from the Tuscaloosa Marine Shale region. Color bar gives the probability of occurrence.}
\label{fig:RfampSim}
\end{figure*}

\subsection{Analytic Model} 

Peak Rayleigh wave amplitude can be approximately modeled as a function in earthquake magnitude, depth and distance \citep{CoEa2017} as follows
\begin{equation}\label{eq:Rfamp}
Rf_{amp} = M \frac{a}{f^{b}_{\mathbb{C}}} \frac{e^{-2\pi h f_{\mathbb{C}} /c}}{r^{d}}\;
\end{equation}
where $M$ is the magnitude of the earthquake, $f_{\mathbb{C}} = 10^{2.3-M/2}\,\rm$ is the corner frequency of the earthquake,  $h$ is the depth, $r$ is the distance to the detector, and $c$ is the speed of the surface-waves, all in SI units.  Here the free parameters \emph{a,b,c,d} are best-fit parameters which make the model predict 90 \% of the observed motion within a factor of five. For this simulation, we use a = 0.16, b = 1.31, c = 4672.83 and d = 0.81. As for magnitude and depth, we assume a distribution similar to the average US induced seismicity data available since 1967.  Within the TMS region, we choose regions with high resistivities (of order seven $ohm \times m$ and higher), as they have a higher chance of being exploited for oil \citep{Amelia_Resitivity}. We randomly sample the parameter distribution and estimate the peak amplitude motion using Eq.~\ref{eq:Rfamp}. Results are shown in Fig.~\ref{fig:RfampSim}, where we observe more than 50\% of the events to cause motion of order tens of microns per second. Even magnitude 2.7 earthquakes this close to the Livingston site would severely impact instrument operation.

\begin{figure*}[!htb]
\includegraphics[width=0.75\textwidth]{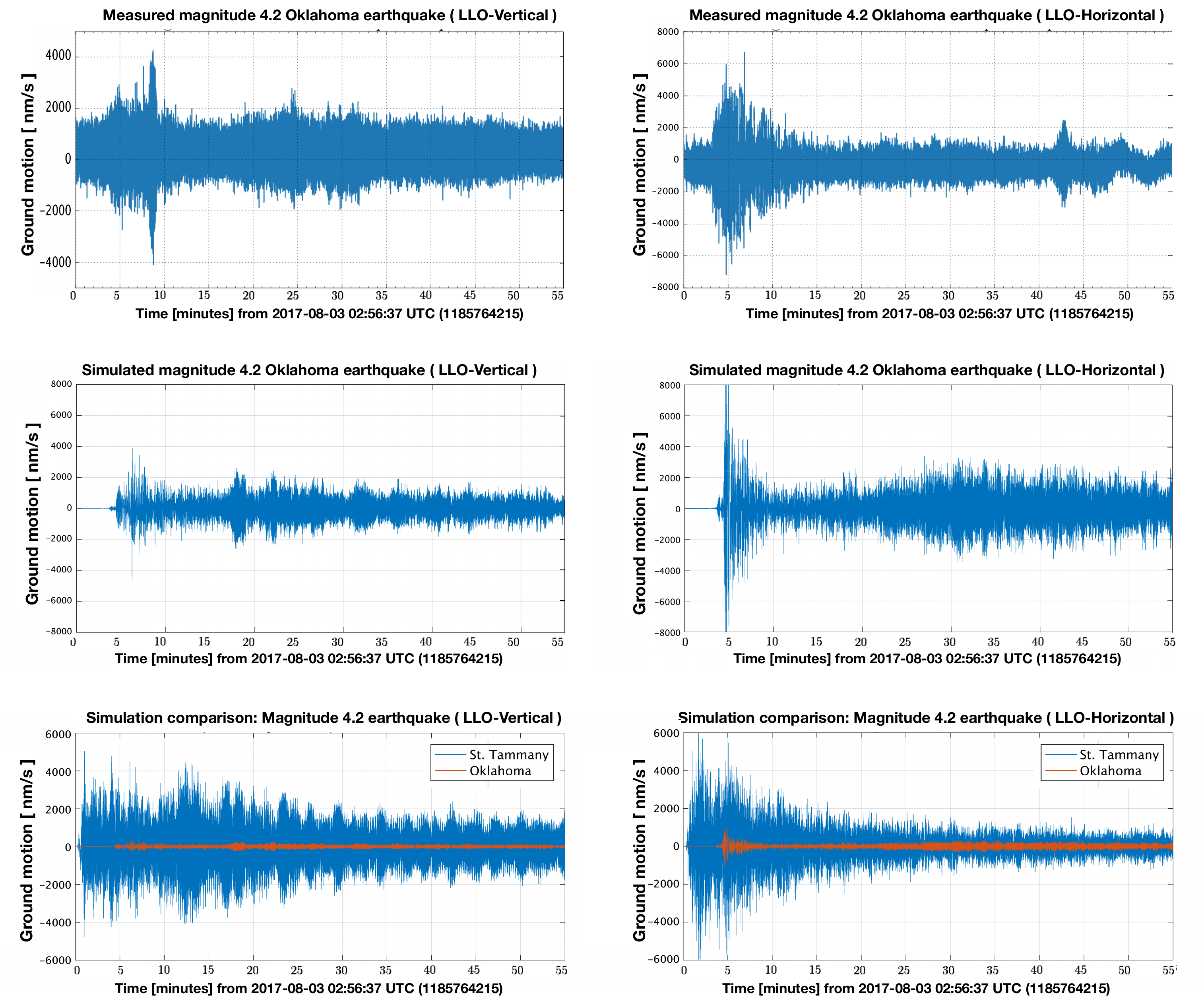}
\caption[Expected ground motion along the vertical and horizontal direction from a 4.2 magnitude earthquake]{\label{fig:RfampSimTMS} Expected ground motion along the vertical and horizontal direction from a 4.2 magnitude earthquake triggered at St. Tammany Parish (Tuscaloosa Marine Shale) and Oklahoma region modeled using SPECFEM3D. The inset plots show the measured ground motion for a similar earthquake that happened on Aug 3, 2018, at Edmond, Oklahoma.}
\end{figure*}

\subsection{Spectral Element Model of the fault rupture}\label{subsec:chD:specfem}

Here we present the results from the forward simulation of the earthquake and compare it with observed seismogram data. The fault rupture simulations are carried out through the SPECFEM3D software \citep{KoVi98,TrKoLi08} through the method of spectral finite elements \citep{patera1984spectral} which is well suited for wave propagation problems that require high accuracy. It is a special case of the Galerkin method \citep{Cockburn:2011:DGM:2408658} where the computational error is minimized through the use of high degree piecewise polynomial basis functions such as Legendre or Chebyshev functions (See Appendix~\ref{App1} for details). We use a half-space with multiple layers with different elastic properties. A hexahedral meshing of the layered half-space is carried out using an internal mesher, xmeshfem3d \citep{komatitsch1998spectral,komatitsch1999introduction}. SCOTCH library\citep{pellegrini:1996:SSP:645560.658570} provides efficient mesh and graph partitioning for load partitioning in HPC systems. This process is succeeded by database generation and finally the spectral finite element solver. We do not include effects from topography in this simulation.

We recreate the 4.2 Oklahoma earthquake occurred on 3rd August 2018 at Edmond, Oklahoma \citep{OkhCMT_USGS} and compare the results with the actual seismometer readings obtained from the LLO site. We then compare the results by shifting the fault rupture location to St. Tammany Parish region, Louisiana.  For the forward simulation, we require centroid moment tensor (CMT) values that describe the fault rupture obtained from the United States Geological Survey (USGS) in addition to the location, depth, and local geological information. Mesh of depth 60 km is used with absorbing boundary conditions. We assume the compressional ($V_{p}$) and shear velocity ($V_{s}$) to be uniform from the source to destination and to follow a vertical trend typical of what is observed in Louisiana. We use results from \citet{gaite20153} to get the shear wave velocity ($V_{s}$) till depths of up to 30 km. $V_{p}$ is calculated using Castagna's relation which provides a model for regions near the Gulf Of Mexico as
\begin{equation}
V_{p} = 1.13 \;  V_{s} + 1360 \;.
\end{equation}
Our model uses results from \citep{castagna1993avo} for the variation of Shale density as a function of depth. Fig.~\ref{fig:RfampSimTMS} shows the results of our simulation. We observe the close similarity between the simulated seismogram and the seismometer reading both regarding arrival time and the peak amplitude. It is evident that a similar event from the TMS region would hit the site in less than a minute with an amplitude of 60 microns/second, an order of magnitude higher than the original Oklahoma event. Such events triggered by the exploration and exploitation of Tuscaloosa Marine Shale would be detrimental to the operation of LIGO Livingston.

\section{Strategies for impact minimization}\label{sec:IMPACT}

Several of the seismic isolation control loops designed at LIGO are optimized for providing noise suppression at frequencies close to and above the microseism leading to extra gain peaking below 30 mHz. This makes the instrument susceptible to noise re-injection during periods of elevated ground motion.  Proximity to regions with a high rate of induced seismicity could be detrimental in maintaining steady lock for current and future generation detectors. Future underground observatory like Einstein Telescope~\citep{abernathy2011einstein} is likely to be built at a depth of $100-200$~m to avoid the effects of surface Rayleigh waves and more specifically the Newtonian Noise~(NN) arising from the local density perturbations caused by the propagating seismic fields \citep{saulson1984terrestrial,hughes1998seismic}. NN is a dominant sensitivity limiting noise at frequencies below 20 Hz and will require a post-offline  subtraction via Wiener filtering using an optimal array of seismic sensors~\citep{Harms2015, coughlin2016towards,coughlin2018implications}. 
A swarm of low magnitude induced earthquakes could lead to a persistent seismic background which in general would be less dominant to the other competing low-frequency noises. But the Rayleigh waves from these could cause density perturbations around the vicinity of the detector and contribute to the already existing Newtonian noise. Microquakes from the induced seismicity ``hot spots'' could also alter the seismic correlation patterns thus affecting the overall subtraction efficiency of the optimal Wiener filter. Thus it is advisable to assess the level of induced seismicity prior to site selection for the next generation underground terrestrial interferometers. In the future, we would like to make use of the spectral finite element methods described in Section~\ref{subsec:chD:specfem} to probe the NN background from induced events both as a function of depth as well as its frequency.

Analysis carried out in the previous section also necessitates the need for very low latency warning systems \citep{kohler2018earthquake} whose inputs would be crucial in deciding the appropriate control strategy that minimizes the seismic impact. Warning systems based on gravity gradiometers could be another option well suited to issue such low latency alerts. Previous works~\citep{harms2015transient,montagner2016prompt} have shown the possibility of detecting transient changes in gravity within 10\; sec after fault rupture. Such extremely sensitive gravity strainmeters will not be limited by seismic propagation speed and may be able to work independently of the seismometers to inform the site commissioners well in advance of the arrival of the P waves.
\begin{figure}[h]
\begin{center}
\includegraphics[width=\linewidth]{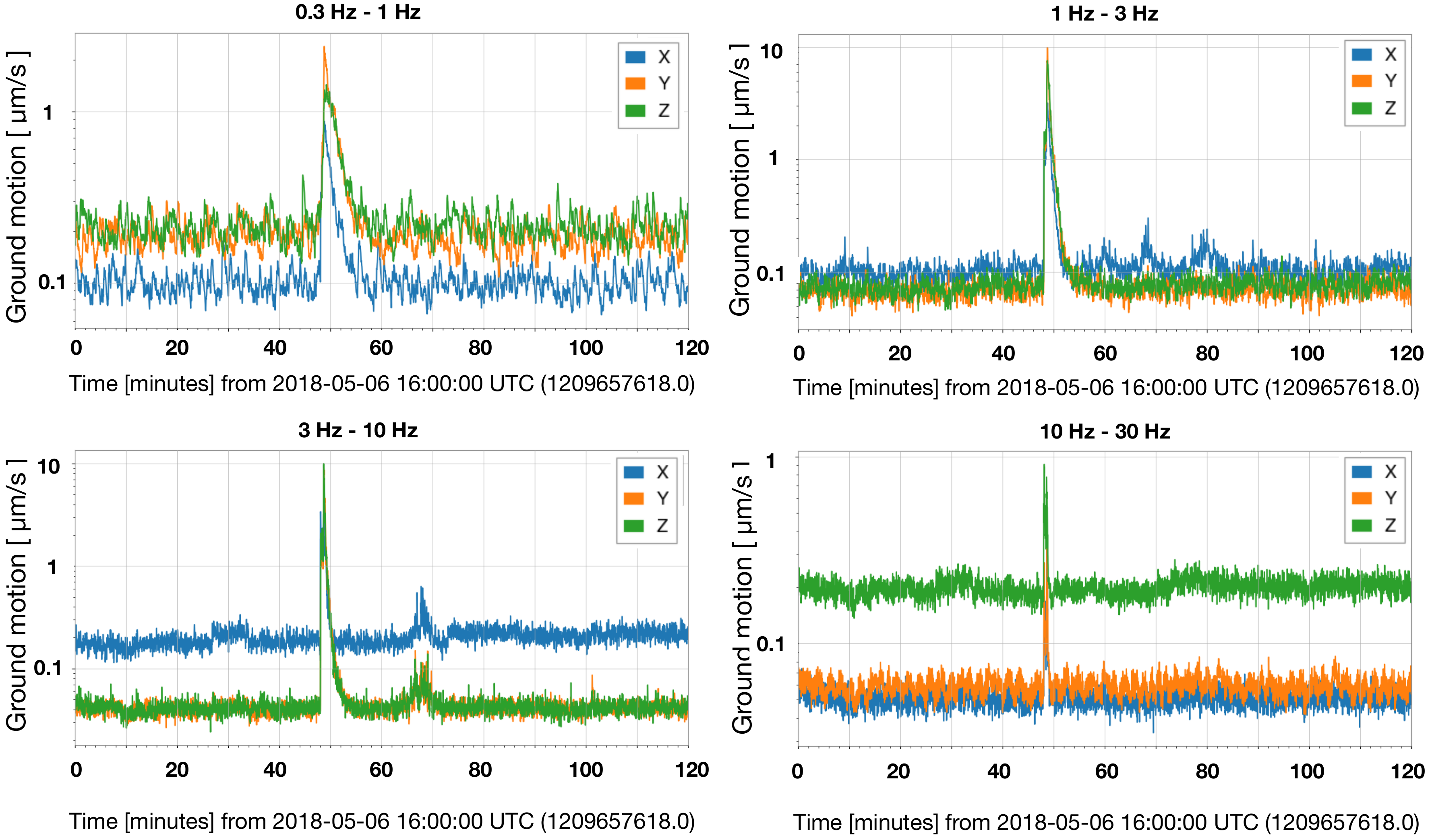}
\caption{Observation of high-frequency content in LLO seismometer data caused by the 4.6M earthquake that occurred on 2018-05-06 16:47:09 UTC at the Gulf of Mexico (182 km SSE of Buras-Triumph, Louisiana at a depth of 10 m ). }
\label{fig:gulf}
\end{center}
\end{figure}
The recent 4.6 magnitude earthquake that occurred in the Gulf of Mexico on 2018-05-06 16:47:09 UTC at a depth of 10 km registered a band limited RMS well above 6 microns along the horizontal direction in $1-10$~Hz and was strongly witnessed in the $10-30$~Hz band (See Fig. \ref{fig:gulf}). We have ruled out the possibility of saturation or up-conversion in the seismometer by cross verifying the signal using geo-phones installed at the same location. Although the event is not linked with induced seismicity, it provides an idea about how geographical nearness and shallowness in depth can affect the frequency bands that are of astrophysical significance for the next generation detectors. This observation strengthens the need for future earthquake isolation designs to also consider frequency bands above the traditional $30-100$~mHz range.

\begin{figure}[!htb]
\begin{center}
\includegraphics[width=\linewidth]{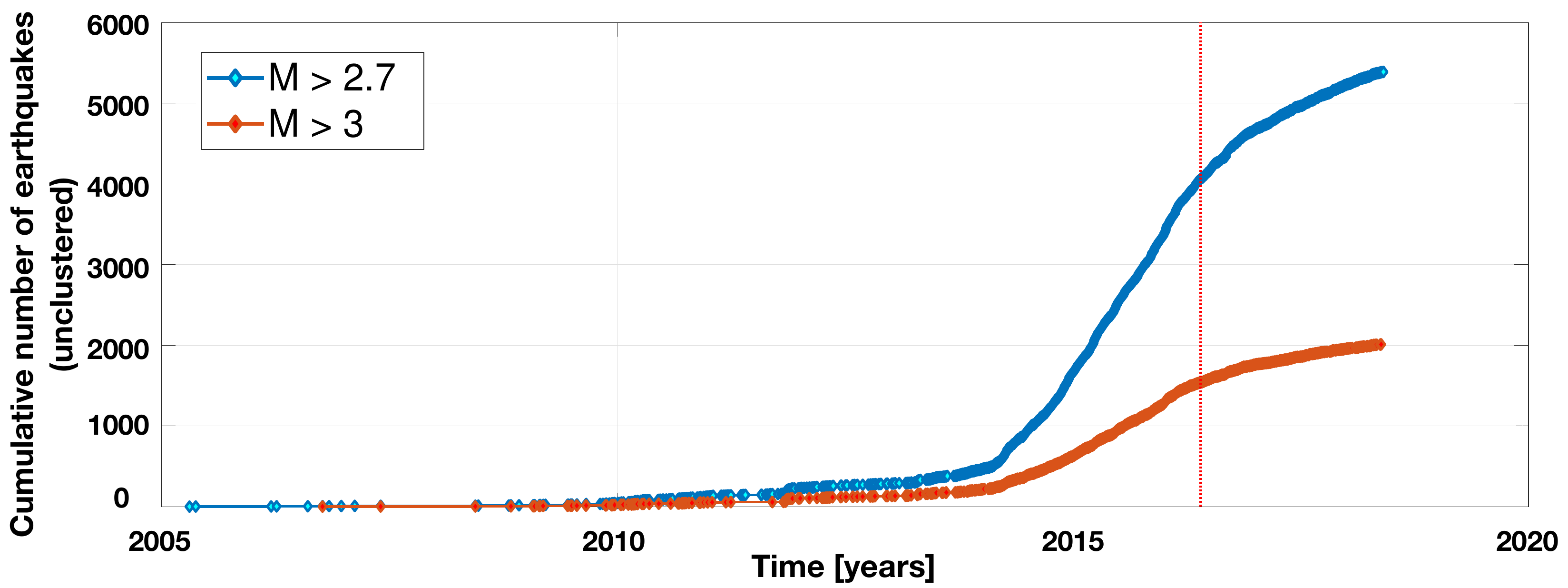}
\caption{The cumulative number of earthquakes with magnitude greater than three that happened in Oklahoma (orange line). Regulations in place since May 28, 2016 (red line), is seen to decrease the frequency of these events.}
\label{fig:OKHregulation} 
\end{center}
\end{figure}

It has been demonstrated that restrictions on drilling newer wells near areas of reported seismic activity and curbing the operations of existing injection wells lead to a decrease in induced earthquakes. For example, the state of Arkansas saw an upsurge of earthquakes in 2009 which dropped since 2011 due to prompt legislative measures \citep{Arkansas_weblink}. The rate of wastewater fluid injection often correlates with the number of observed earthquakes as is seen from the Rocky Mountain Arsenal data~\citep{osti_6805542,hsieh1981reservoir}. Controlling the amount of injected fluid could be one strategy to minimize the triggering. Restrictions imposed at Oklahoma on the rate of fluid injections since May 28, 2016  \citep{Oklahoma_restrictions} led to a decrease in the number of earthquakes and is seen in Fig.~\ref{fig:OKHregulation} which shows the cumulative number of earthquakes with magnitude greater than three that occurred in the state.

\section{Conclusions} \label{CONCLUSION}
In this paper, we have discussed the effects of induced seismicity on gravitational wave interferometers. In particular, we assessed the impact of events from Oklahoma and showed that due to the geographical nearness and the shallowness of the hypo-center, earthquakes of magnitude as small as 4.2 from these areas could lead to lockloss scenarios at the Livingston detector. The closeness of the source also results in a contribution to the spectral content beyond the earthquake band which can affect the seismic control loops deployed at the interferometer. 

We further estimated the possible effects to LIGO Livingston detector from the Tuscaloosa Marine Shale formation which is a potential candidate for future oil extraction in the conterminous US. Proximity to the site implies even minor quakes from these regions could be a potential cause of lockloss at the interferometers. The latency of early warning system currently installed at the site is around 20 minutes which will not be sufficient to handle events described here. Alternative schemes based on gravity gradiometers need to be explored for prompt early detection and warning.  Accessing the induced seismicity near the site will be beneficial for the operation of next-generation ground-based and underground GW observatories. This would require extensive monitoring of low magnitude events and subsequent classification based on their peculiar features. A better understanding of the underlying fault structure coupled with a reduced rate of injection at class II wells and ensuring that the wells are located away from the stressed faults could minimize the impact of induced seismicity on the ground-based interferometers.

\section{Acknowledgments} \label{ACKNOWLEDGEMENTS}
We thank the seismic \& suspensions working group and the detector characterization group of the LIGO Scientific Collaboration for their comments and suggestions. N.M. acknowledges the Council for Scientific and Industrial Research (CSIR), India for providing financial support as a Senior Research Fellow.  S. M. acknowledges support from the Department of Science \& Technology (DST), India provided under the Swarna Jayanti Fellowships scheme. This research benefited from a grant awarded to IUCAA by the  Navajbai Ratan Tata Trust (NRTT). The authors express thanks to Juan Lorenzo, Rana Adhikari, Sharad Gaonkar and Arnaud Pele for their valuable comments and suggestions. Simulation results shown in this work were carried out using IUCAA Perseus cluster. LIGO was constructed by the California Institute of Technology and Massachusetts Institute of Technology with funding from the National Science Foundation and operates under Cooperative Agreement No.PHY-0757058. This paper has been assigned LIGO Document No. LIGO-P1800289.

\appendix 

\section{}\label{App1}
We briefly describe below the spectral finite element scheme used for the forward simulation of the earthquake.  In terms of displacement vector $\boldsymbol{u}$, Cauchy strain tensor $\boldsymbol{\sigma}$ and seismic excitation vector $\boldsymbol{f}$, the elastic wave equation in a continuous medium has the following form,

\begin{equation}\label{EQ:elasticWave}
\rho \; \boldsymbol{\ddot  u} - \nabla \cdot \boldsymbol{\sigma} = \boldsymbol{f} \;.
\end{equation}

A weak or variational formulation of Eq. \ref{EQ:elasticWave} is obtained by taking dot product with a continuous and well behaved test vector ($\boldsymbol{v}$) followed by integration by parts with the appropriate boundary conditions which leads to,

\begin{equation}\label{EQ:weakForm}
\int_{G}\rho \boldsymbol{v}\cdot \boldsymbol{\ddot{u}} \;d^{3}x + \int_{G} \nabla \boldsymbol{v} . \boldsymbol{\sigma} \;d^{3}x  = \int_{G}\boldsymbol{v} \cdot \boldsymbol{f} \; d^{3}x \;.
\end{equation}
Standard strategy is to find those solutions which satisfy the weak formulation but at the same time are sufficiently continuous and differentiable so as to satisfy the strong form. Solution to Eq. \ref{EQ:weakForm} is obtained by discretizing the spatial domain into smaller subdomains $G_{e}$ which are mapped on to a reference cube $\mathcal{F}_{e}:\small{[}-1,1\small{]}^{3} \rightarrow G_{e} \text{ such that }\boldsymbol{x}(\zeta) = \mathcal{F}_{e}(\zeta)$. Weak formulation can be solved for each subdomain as,

\begin{equation}\label{EQ:weakFormIndividual}
\resizebox{.8\hsize}{!}{$
\int_{-1}^{1}\rho \boldsymbol{v}\cdot \boldsymbol{\ddot{u}} \;\boldsymbol{J}(\boldsymbol{\zeta})\;d^{3}\boldsymbol{\zeta} + \int_{-1}^{1} \nabla \boldsymbol{v} . \boldsymbol{\sigma}\;\boldsymbol{J}(\zeta)\;d^{3}\boldsymbol{\zeta} =  \int_{-1}^{1}\boldsymbol{v} \cdot \boldsymbol{f} \; \boldsymbol{J}(\zeta)\;d^{3}\boldsymbol{\zeta} \;,
$}
\end{equation}
where J is the Jacobian of $\mathcal{F}_{e}$. Langrange polynomial ($\ell^{N}$) based interpolation is used to project the displacement field on to the reference grid,

\begin{equation}
\boldsymbol{u}_{p}(\boldsymbol{x},t)_{G_{e}} = 
\sum_{i,j,k=1}^{N+1}\; u_{p}^{ijk}(t)\psi_{ijk}(\boldsymbol{x}) \;,
\end{equation}
with 
\begin{equation}
\psi_{ijk}(\boldsymbol{x(\zeta)}) = \ell_{i}^{N}(\zeta_{1})\;\ell_{j}^{N}(\zeta_{2})\;\ell_{k}^{N}(\zeta_{3}), \; \zeta \in [-1,1] \;.
\end{equation}

Each of the integrals is calculated using Gauss-Lobatto-Legendre-Birkhoff (GLLB) quadrature \citep{hesthaven2007spectral}.  Thus for every subdomain element we obtain two matrices: an elemental mass matrix, 
\begin{equation}
M^{e}_{\alpha \beta} = \rho_{\beta}\omega_{\beta}\mathcal{J}_{\beta}\delta_{\alpha \beta}\; ,
\end{equation}
expressed in terms of density $\rho$, determinant of Jacobi matrix $\mathcal{J}$ and weights associated with GLLB quadrature of integration; 
and a stiffness matrix,
\begin{equation}
K_{ij}^{e} = \sum_{i=0}^{N} \sum_{k=0}^{N}  \mu_{k}\; \omega_{k} \;^{i}J^{2}_{k}\; \mathcal{J}_{k}\; \ell_{i}^{'}(\zeta_{k}) \;\ell_{j}^{'}(\zeta_{k}) \; ,
\end{equation}
where $^{i}J$ is the Jacobi matrix corresponding to the coordinate transformation from global to local coordinates. GLLB quadrature coupled with the use of Lagrange polynomials leads to an exactly diagonal mass matrix which is easy to invert and has lower memory requirements. It also results in a simple explicit time scheme which can take advantage of parallel processing infrastructure. The assembling the matrices ($\boldsymbol{M}^{e}$ \& $\boldsymbol{K}^{e}$ ) as well as the forces ($\boldsymbol{u}^{e}$ \& $\boldsymbol{f}^{e}$) from each individual element results in the following global matrix equation,
\begin{equation}
\boldsymbol{M}  \ddot{\boldsymbol{u}}(t) + \boldsymbol{K}  \boldsymbol{u}(t) = \boldsymbol{f}(t) \;.
\end{equation}
The displacement $\boldsymbol{u}$ at the $t + dt$ interval can then be obtained by a centered finite-difference approximation of the second derivative,
\begin{equation}
\boldsymbol{u}(t+dt) = dt^{2}\Big[ \boldsymbol{M}^{-1} \Big(\boldsymbol{f}(t) - \boldsymbol{K}\;\boldsymbol{u}(t) \Big) \Big]  + 2\boldsymbol{u}(t) - \boldsymbol{u}(t-dt) \; .
\end{equation}

\section{}\label{App2}

\begin{table*}
\resizebox{\linewidth}{!}
{
\begin{tabular}{lllrrrr}
\toprule
{} &time & place &  mag &  latitude &  longitude &  depth \\
\midrule
0  &  2018-04-09T10:22:20.050Z &    27km WSW of Perry, Oklahoma &  4.5 &   36.2165 &   -97.5820 &  3.890 \\
1  &  2018-04-07T12:16:03.300Z &      20km W of Perry, Oklahoma &  4.6 &   36.2896 &   -97.5166 &  5.791 \\
2  &  2018-03-05T03:40:59.200Z &      15km NE of Enid, Oklahoma &  4.2 &   36.4833 &   -97.7491 &  8.229 \\
3  &  2018-03-04T23:17:17.400Z &      15km NE of Enid, Oklahoma &  4.2 &   36.4780 &   -97.7411 &  8.103 \\
4  &  2017-09-08T02:26:23.700Z &  12km SSE of Medford, Oklahoma &  4.3 &   36.6996 &   -97.6830 &  6.073 \\
5  &  2017-08-03T02:56:37.300Z &    6km ENE of Edmond, Oklahoma &  4.2 &   35.6720 &   -97.4047 &  5.031 \\
6  &  2017-07-14T13:47:35.300Z &     12km N of Stroud, Oklahoma &  4.2 &   35.8591 &   -96.6833 &  6.813 \\
7  &  2017-05-13T08:32:37.300Z &  32km NW of Fairview, Oklahoma &  4.2 &   36.4782 &   -98.7365 &  5.367 \\
8  &  2016-11-07T01:44:24.500Z &     3km W of Cushing, Oklahoma &  5.0 &   35.9907 &   -96.8030 &  4.430 \\
9  &  2016-11-02T04:26:54.000Z &   12km ESE of Pawnee, Oklahoma &  4.4 &   36.3047 &   -96.6657 &  4.349 \\
10 &  2016-09-03T12:02:44.400Z &    14km NW of Pawnee, Oklahoma &  5.8 &   36.4251 &   -96.9291 &  5.557 \\
11 &  2016-07-17T04:17:58.900Z &      20km W of Perry, Oklahoma &  4.2 &   36.2841 &   -97.5138 &  4.787 \\
12 &  2016-07-09T02:04:27.400Z &  33km NW of Fairview, Oklahoma &  4.4 &   36.4638 &   -98.7584 &  7.242 \\
13 &  2016-07-08T22:29:38.000Z &  33km NW of Fairview, Oklahoma &  4.2 &   36.4749 &   -98.7459 &  6.356 \\
14 &  2016-07-08T21:31:57.600Z &  32km NW of Fairview, Oklahoma &  4.2 &   36.4765 &   -98.7387 &  7.315 \\
15 &  2016-04-07T22:27:30.400Z &      1km E of Luther, Oklahoma &  4.2 &   35.6619 &   -97.1741 &  6.106 \\
16 &  2016-03-29T04:53:01.200Z &  4km NNE of Crescent, Oklahoma &  4.2 &   35.9900 &   -97.5773 &  5.183 \\
17 &  2016-02-13T17:07:06.290Z &  31km NW of Fairview, Oklahoma &  5.1 &   36.4898 &   -98.7090 &  8.310 \\
18 &  2016-01-07T08:37:11.100Z &  32km NW of Fairview, Oklahoma &  4.4 &   36.4754 &   -98.7342 &  6.647 \\
19 &  2016-01-07T04:27:57.600Z &  33km NW of Fairview, Oklahoma &  4.7 &   36.4955 &   -98.7254 &  4.058 \\
20 &  2016-01-07T04:27:27.900Z &  33km NW of Fairview, Oklahoma &  4.4 &   36.4860 &   -98.7412 &  7.093 \\
21 &  2016-01-01T11:39:39.800Z &    6km ENE of Edmond, Oklahoma &  4.2 &   35.6688 &   -97.4065 &  5.825 \\
22 &  2015-12-29T11:39:19.200Z &    6km ENE of Edmond, Oklahoma &  4.3 &   35.6654 &   -97.4054 &  6.532 \\
23 &  2015-11-30T09:49:12.800Z &   26km E of Cherokee, Oklahoma &  4.7 &   36.7509 &   -98.0561 &  5.629 \\
24 &  2015-11-23T21:17:46.500Z &  11km NE of Cherokee, Oklahoma &  4.4 &   36.8382 &   -98.2762 &  5.028 \\
25 &  2015-11-19T07:42:12.000Z &  13km SW of Cherokee, Oklahoma &  4.7 &   36.6602 &   -98.4594 &  5.914 \\
26 &  2015-11-15T09:45:31.300Z &  33km NW of Fairview, Oklahoma &  4.3 &   36.4696 &   -98.7549 &  5.109 \\
27 &  2015-10-10T22:03:05.300Z &     3km W of Cushing, Oklahoma &  4.3 &   35.9860 &   -96.8032 &  3.274 \\
28 &  2015-10-10T09:20:43.000Z &  20km WSW of Medford, Oklahoma &  4.4 &   36.7187 &   -97.9311 &  5.629 \\
29 &  2015-07-27T18:12:15.400Z &  4km NNE of Crescent, Oklahoma &  4.5 &   35.9889 &   -97.5717 &  5.000 \\
30 &  2015-07-20T20:19:03.400Z &  13km NE of Cherokee, Oklahoma &  4.4 &   36.8420 &   -98.2593 &  4.079 \\
\bottomrule
\end{tabular}
}
\caption{List of earthquakes that occurred in Oklahoma during the period 2015 - 2018 with a magnitude of at least 4.2 }
\label{table:Oklahoma_events}
\end{table*}

\bibliography{reference.bib}

\begin{thebibliography}{51}
\expandafter\ifx\csname natexlab\endcsname\relax\def\natexlab#1{#1}\fi
\expandafter\ifx\csname bibnamefont\endcsname\relax
  \def\bibnamefont#1{#1}\fi
\expandafter\ifx\csname bibfnamefont\endcsname\relax
  \def\bibfnamefont#1{#1}\fi
\expandafter\ifx\csname citenamefont\endcsname\relax
  \def\citenamefont#1{#1}\fi
\expandafter\ifx\csname url\endcsname\relax
  \def\url#1{\texttt{#1}}\fi
\expandafter\ifx\csname urlprefix\endcsname\relax\def\urlprefix{URL }\fi
\providecommand{\bibinfo}[2]{#2}
\providecommand{\eprint}[2][]{\url{#2}}

\bibitem[{\citenamefont{{Abbott, B. P. et al}}(2016)}]{AbEA2016a}
\bibinfo{author}{\bibnamefont{{Abbott, B. P. et al}}}
  (\bibinfo{collaboration}{LIGO Scientific Collaboration and Virgo
  Collaboration}), \bibinfo{journal}{Phys. Rev. Lett.}
  \textbf{\bibinfo{volume}{116}}, \bibinfo{pages}{061102}
  (\bibinfo{year}{2016}),
  \urlprefix\url{http://link.aps.org/doi/10.1103/PhysRevLett.116.061102}.

\bibitem[{\citenamefont{{Abbott, B. P. et al.}}(2016)}]{AbEA2016e}
\bibinfo{author}{\bibnamefont{{Abbott, B. P. et al.}}}
  (\bibinfo{collaboration}{LIGO Scientific Collaboration and Virgo
  Collaboration}), \bibinfo{journal}{Phys. Rev. Lett.}
  \textbf{\bibinfo{volume}{116}}, \bibinfo{pages}{241103}
  (\bibinfo{year}{2016}),
  \urlprefix\url{http://link.aps.org/doi/10.1103/PhysRevLett.116.241103}.

\bibitem[{\citenamefont{{Abbott, B. P. et
  al.}}(2017{\natexlab{a}})}]{AbEA2017a}
\bibinfo{author}{\bibnamefont{{Abbott, B. P. et al.}}}
  (\bibinfo{collaboration}{LIGO Scientific and Virgo Collaboration}),
  \bibinfo{journal}{Phys. Rev. Lett.} \textbf{\bibinfo{volume}{118}},
  \bibinfo{pages}{221101} (\bibinfo{year}{2017}{\natexlab{a}}),
  \urlprefix\url{https://link.aps.org/doi/10.1103/PhysRevLett.118.221101}.

\bibitem[{\citenamefont{{Abbott, B. P. et
  al.}}(2017{\natexlab{b}})}]{AbEA2017b}
\bibinfo{author}{\bibnamefont{{Abbott, B. P. et al.}}}
  (\bibinfo{collaboration}{LIGO Scientific Collaboration and Virgo
  Collaboration}), \bibinfo{journal}{Phys. Rev. Lett.}
  \textbf{\bibinfo{volume}{119}}, \bibinfo{pages}{161101}
  (\bibinfo{year}{2017}{\natexlab{b}}),
  \urlprefix\url{https://link.aps.org/doi/10.1103/PhysRevLett.119.161101}.

\bibitem[{\citenamefont{{Abbott, B. P. et
  al.}}(2017{\natexlab{c}})}]{AbEA2017c}
\bibinfo{author}{\bibnamefont{{Abbott, B. P. et al.}}}
  (\bibinfo{collaboration}{LIGO Scientific Collaboration and Virgo
  Collaboration}), \bibinfo{journal}{Phys. Rev. Lett.}
  \textbf{\bibinfo{volume}{119}}, \bibinfo{pages}{141101}
  (\bibinfo{year}{2017}{\natexlab{c}}),
  \urlprefix\url{https://link.aps.org/doi/10.1103/PhysRevLett.119.141101}.

\bibitem[{\citenamefont{Matichard et~al.}(2015)\citenamefont{Matichard, Lantz,
  Mittleman, Mason, Kissel, Abbott, Biscans, McIver, Abbott, Abbott
  et~al.}}]{MaLa2015}
\bibinfo{author}{\bibfnamefont{F.}~\bibnamefont{Matichard}},
  \bibinfo{author}{\bibfnamefont{B.}~\bibnamefont{Lantz}},
  \bibinfo{author}{\bibfnamefont{R.}~\bibnamefont{Mittleman}},
  \bibinfo{author}{\bibfnamefont{K.}~\bibnamefont{Mason}},
  \bibinfo{author}{\bibfnamefont{J.}~\bibnamefont{Kissel}},
  \bibinfo{author}{\bibfnamefont{B.}~\bibnamefont{Abbott}},
  \bibinfo{author}{\bibfnamefont{S.}~\bibnamefont{Biscans}},
  \bibinfo{author}{\bibfnamefont{J.}~\bibnamefont{McIver}},
  \bibinfo{author}{\bibfnamefont{R.}~\bibnamefont{Abbott}},
  \bibinfo{author}{\bibfnamefont{S.}~\bibnamefont{Abbott}},
  \bibnamefont{et~al.}, \bibinfo{journal}{Classical and Quantum Gravity}
  \textbf{\bibinfo{volume}{32}}, \bibinfo{pages}{185003}
  (\bibinfo{year}{2015}),
  \urlprefix\url{http://stacks.iop.org/0264-9381/32/i=18/a=185003}.

\bibitem[{\citenamefont{Biscans et~al.}(2018)\citenamefont{Biscans, Warner,
  Mittleman, Buchanan, Coughlin, Evans, Gabbard, Harms, Lantz, Mukund
  et~al.}}]{BiWa2018}
\bibinfo{author}{\bibfnamefont{S.}~\bibnamefont{Biscans}},
  \bibinfo{author}{\bibfnamefont{J.}~\bibnamefont{Warner}},
  \bibinfo{author}{\bibfnamefont{R.}~\bibnamefont{Mittleman}},
  \bibinfo{author}{\bibfnamefont{C.}~\bibnamefont{Buchanan}},
  \bibinfo{author}{\bibfnamefont{M.}~\bibnamefont{Coughlin}},
  \bibinfo{author}{\bibfnamefont{M.}~\bibnamefont{Evans}},
  \bibinfo{author}{\bibfnamefont{H.}~\bibnamefont{Gabbard}},
  \bibinfo{author}{\bibfnamefont{J.}~\bibnamefont{Harms}},
  \bibinfo{author}{\bibfnamefont{B.}~\bibnamefont{Lantz}},
  \bibinfo{author}{\bibfnamefont{N.}~\bibnamefont{Mukund}},
  \bibnamefont{et~al.}, \bibinfo{journal}{Classical and Quantum Gravity}
  \textbf{\bibinfo{volume}{35}}, \bibinfo{pages}{055004}
  (\bibinfo{year}{2018}),
  \urlprefix\url{https://doi.org/10.1088%2F1361-6382%2Faaa4aa}.

\bibitem[{\citenamefont{Coughlin
  et~al.}(2017{\natexlab{a}})\citenamefont{Coughlin, Earle, Harms, Biscans,
  Buchanan, Coughlin, Donovan, Fee, Gabbard, Guy et~al.}}]{seismon1}
\bibinfo{author}{\bibfnamefont{M.}~\bibnamefont{Coughlin}},
  \bibinfo{author}{\bibfnamefont{P.}~\bibnamefont{Earle}},
  \bibinfo{author}{\bibfnamefont{J.}~\bibnamefont{Harms}},
  \bibinfo{author}{\bibfnamefont{S.}~\bibnamefont{Biscans}},
  \bibinfo{author}{\bibfnamefont{C.}~\bibnamefont{Buchanan}},
  \bibinfo{author}{\bibfnamefont{E.}~\bibnamefont{Coughlin}},
  \bibinfo{author}{\bibfnamefont{F.}~\bibnamefont{Donovan}},
  \bibinfo{author}{\bibfnamefont{J.}~\bibnamefont{Fee}},
  \bibinfo{author}{\bibfnamefont{H.}~\bibnamefont{Gabbard}},
  \bibinfo{author}{\bibfnamefont{M.}~\bibnamefont{Guy}}, \bibnamefont{et~al.},
  \bibinfo{journal}{Classical and Quantum Gravity}
  \textbf{\bibinfo{volume}{34}}, \bibinfo{pages}{044004}
  (\bibinfo{year}{2017}{\natexlab{a}}),
  \urlprefix\url{http://stacks.iop.org/0264-9381/34/i=4/a=044004}.

\bibitem[{\citenamefont{Mukund et~al.}(2019)\citenamefont{Mukund, Coughlin,
  Harms, Biscans, Warner, Pele, Thorne, Barker, Arnaud, Donovan
  et~al.}}]{seismon2}
\bibinfo{author}{\bibfnamefont{N.}~\bibnamefont{Mukund}},
  \bibinfo{author}{\bibfnamefont{M.}~\bibnamefont{Coughlin}},
  \bibinfo{author}{\bibfnamefont{J.}~\bibnamefont{Harms}},
  \bibinfo{author}{\bibfnamefont{S.}~\bibnamefont{Biscans}},
  \bibinfo{author}{\bibfnamefont{J.}~\bibnamefont{Warner}},
  \bibinfo{author}{\bibfnamefont{A.}~\bibnamefont{Pele}},
  \bibinfo{author}{\bibfnamefont{K.}~\bibnamefont{Thorne}},
  \bibinfo{author}{\bibfnamefont{D.}~\bibnamefont{Barker}},
  \bibinfo{author}{\bibfnamefont{N.}~\bibnamefont{Arnaud}},
  \bibinfo{author}{\bibfnamefont{F.}~\bibnamefont{Donovan}},
  \bibnamefont{et~al.}, \bibinfo{journal}{Classical and Quantum Gravity}
  (\bibinfo{year}{2019}),
  \urlprefix\url{http://iopscience.iop.org/10.1088/1361-6382/ab0d2c}.

\bibitem[{\citenamefont{Petersen et~al.}(2018)\citenamefont{Petersen, Mueller,
  Moschetti, Hoover, Rukstales, McNamara, Williams, Shumway, Powers, Earle
  et~al.}}]{induced_seismicity_catalog}
\bibinfo{author}{\bibfnamefont{M.}~\bibnamefont{Petersen}},
  \bibinfo{author}{\bibfnamefont{C.}~\bibnamefont{Mueller}},
  \bibinfo{author}{\bibfnamefont{M.}~\bibnamefont{Moschetti}},
  \bibinfo{author}{\bibfnamefont{S.}~\bibnamefont{Hoover}},
  \bibinfo{author}{\bibfnamefont{K.}~\bibnamefont{Rukstales}},
  \bibinfo{author}{\bibfnamefont{D.}~\bibnamefont{McNamara}},
  \bibinfo{author}{\bibfnamefont{R.}~\bibnamefont{Williams}},
  \bibinfo{author}{\bibfnamefont{A.}~\bibnamefont{Shumway}},
  \bibinfo{author}{\bibfnamefont{P.}~\bibnamefont{Powers}},
  \bibinfo{author}{\bibfnamefont{P.}~\bibnamefont{Earle}},
  \bibnamefont{et~al.}, \emph{\bibinfo{title}{Seismicity catalog used in the
  2018 one-year seismic hazard forecast for the central and eastern united
  states from induced and natural earthquakes}} (\bibinfo{year}{2018}),
  \urlprefix\url{https://doi.org/10.5066/F7Cf9PC4}.

\bibitem[{\citenamefont{Scholz}(2002)}]{scholz2002mechanics}
\bibinfo{author}{\bibfnamefont{C.~H.} \bibnamefont{Scholz}},
  \emph{\bibinfo{title}{The mechanics of earthquakes and faulting}}
  (\bibinfo{publisher}{Cambridge university press}, \bibinfo{year}{2002}).

\bibitem[{\citenamefont{Zoback}(2010)}]{zoback2010reservoir}
\bibinfo{author}{\bibfnamefont{M.~D.} \bibnamefont{Zoback}},
  \emph{\bibinfo{title}{Reservoir geomechanics}} (\bibinfo{publisher}{Cambridge
  University Press}, \bibinfo{year}{2010}).

\bibitem[{\citenamefont{Council et~al.}(2013)}]{national2013induced}
\bibinfo{author}{\bibfnamefont{N.~R.} \bibnamefont{Council}}
  \bibnamefont{et~al.}, \emph{\bibinfo{title}{Induced seismicity potential in
  energy technologies}} (\bibinfo{publisher}{National Academies Press},
  \bibinfo{year}{2013}).

\bibitem[{\citenamefont{Aki and Richards}(2002)}]{aki2002quantitative}
\bibinfo{author}{\bibfnamefont{K.}~\bibnamefont{Aki}} \bibnamefont{and}
  \bibinfo{author}{\bibfnamefont{P.~G.} \bibnamefont{Richards}},
  \emph{\bibinfo{title}{Quantitative seismology}}
  (\bibinfo{publisher}{University Science Books; Second edition (March 25,
  2009)}, \bibinfo{year}{2002}).

\bibitem[{\citenamefont{Hough}(2014)}]{hough2014shaking}
\bibinfo{author}{\bibfnamefont{S.~E.} \bibnamefont{Hough}},
  \bibinfo{journal}{Bulletin of the Seismological Society of America}
  \textbf{\bibinfo{volume}{104}}, \bibinfo{pages}{2619} (\bibinfo{year}{2014}).

\bibitem[{\citenamefont{Ellsworth et~al.}(2015)\citenamefont{Ellsworth, Llenos,
  McGarr, Michael, Rubinstein, Mueller, Petersen, and
  Calais}}]{ellsworth2015increasing}
\bibinfo{author}{\bibfnamefont{W.~L.} \bibnamefont{Ellsworth}},
  \bibinfo{author}{\bibfnamefont{A.~L.} \bibnamefont{Llenos}},
  \bibinfo{author}{\bibfnamefont{A.~F.} \bibnamefont{McGarr}},
  \bibinfo{author}{\bibfnamefont{A.~J.} \bibnamefont{Michael}},
  \bibinfo{author}{\bibfnamefont{J.~L.} \bibnamefont{Rubinstein}},
  \bibinfo{author}{\bibfnamefont{C.~S.} \bibnamefont{Mueller}},
  \bibinfo{author}{\bibfnamefont{M.~D.} \bibnamefont{Petersen}},
  \bibnamefont{and} \bibinfo{author}{\bibfnamefont{E.}~\bibnamefont{Calais}},
  \bibinfo{journal}{The Leading Edge} \textbf{\bibinfo{volume}{34}},
  \bibinfo{pages}{618} (\bibinfo{year}{2015}).

\bibitem[{\citenamefont{Weingarten et~al.}(2015)\citenamefont{Weingarten, Ge,
  Godt, Bekins, and Rubinstein}}]{weingarten2015high}
\bibinfo{author}{\bibfnamefont{M.}~\bibnamefont{Weingarten}},
  \bibinfo{author}{\bibfnamefont{S.}~\bibnamefont{Ge}},
  \bibinfo{author}{\bibfnamefont{J.~W.} \bibnamefont{Godt}},
  \bibinfo{author}{\bibfnamefont{B.~A.} \bibnamefont{Bekins}},
  \bibnamefont{and} \bibinfo{author}{\bibfnamefont{J.~L.}
  \bibnamefont{Rubinstein}}, \bibinfo{journal}{Science}
  \textbf{\bibinfo{volume}{348}}, \bibinfo{pages}{1336} (\bibinfo{year}{2015}).

\bibitem[{\citenamefont{Keranen et~al.}(2014)\citenamefont{Keranen, Weingarten,
  Abers, Bekins, and Ge}}]{keranen2014sharp}
\bibinfo{author}{\bibfnamefont{K.~M.} \bibnamefont{Keranen}},
  \bibinfo{author}{\bibfnamefont{M.}~\bibnamefont{Weingarten}},
  \bibinfo{author}{\bibfnamefont{G.~A.} \bibnamefont{Abers}},
  \bibinfo{author}{\bibfnamefont{B.~A.} \bibnamefont{Bekins}},
  \bibnamefont{and} \bibinfo{author}{\bibfnamefont{S.}~\bibnamefont{Ge}},
  \bibinfo{journal}{Science} \textbf{\bibinfo{volume}{345}},
  \bibinfo{pages}{448} (\bibinfo{year}{2014}).

\bibitem[{\citenamefont{McNamara
  et~al.}(2015{\natexlab{a}})\citenamefont{McNamara, Benz, Herrmann, Bergman,
  Earle, Holland, Baldwin, and Gassner}}]{mcnamara2015earthquake}
\bibinfo{author}{\bibfnamefont{D.~E.} \bibnamefont{McNamara}},
  \bibinfo{author}{\bibfnamefont{H.~M.} \bibnamefont{Benz}},
  \bibinfo{author}{\bibfnamefont{R.~B.} \bibnamefont{Herrmann}},
  \bibinfo{author}{\bibfnamefont{E.~A.} \bibnamefont{Bergman}},
  \bibinfo{author}{\bibfnamefont{P.}~\bibnamefont{Earle}},
  \bibinfo{author}{\bibfnamefont{A.}~\bibnamefont{Holland}},
  \bibinfo{author}{\bibfnamefont{R.}~\bibnamefont{Baldwin}}, \bibnamefont{and}
  \bibinfo{author}{\bibfnamefont{A.}~\bibnamefont{Gassner}},
  \bibinfo{journal}{Geophysical Research Letters}
  \textbf{\bibinfo{volume}{42}}, \bibinfo{pages}{2742}
  (\bibinfo{year}{2015}{\natexlab{a}}).

\bibitem[{\citenamefont{McNamara
  et~al.}(2015{\natexlab{b}})\citenamefont{McNamara, Hayes, Benz, Williams,
  McMahon, Aster, Holland, Sickbert, Herrmann, Briggs
  et~al.}}]{mcnamara2015reactivated}
\bibinfo{author}{\bibfnamefont{D.~E.} \bibnamefont{McNamara}},
  \bibinfo{author}{\bibfnamefont{G.}~\bibnamefont{Hayes}},
  \bibinfo{author}{\bibfnamefont{H.~M.} \bibnamefont{Benz}},
  \bibinfo{author}{\bibfnamefont{R.}~\bibnamefont{Williams}},
  \bibinfo{author}{\bibfnamefont{N.~D.} \bibnamefont{McMahon}},
  \bibinfo{author}{\bibfnamefont{R.}~\bibnamefont{Aster}},
  \bibinfo{author}{\bibfnamefont{A.}~\bibnamefont{Holland}},
  \bibinfo{author}{\bibfnamefont{T.}~\bibnamefont{Sickbert}},
  \bibinfo{author}{\bibfnamefont{R.}~\bibnamefont{Herrmann}},
  \bibinfo{author}{\bibfnamefont{R.}~\bibnamefont{Briggs}},
  \bibnamefont{et~al.}, \bibinfo{journal}{Geophysical Research Letters}
  \textbf{\bibinfo{volume}{42}}, \bibinfo{pages}{8328}
  (\bibinfo{year}{2015}{\natexlab{b}}).

\bibitem[{\citenamefont{Staley et~al.}(2014)\citenamefont{Staley, Martynov,
  Abbott, Adhikari, Arai, Ballmer, Barsotti, Brooks, DeRosa, Dwyer
  et~al.}}]{0264-9381-31-24-245010}
\bibinfo{author}{\bibfnamefont{A.}~\bibnamefont{Staley}},
  \bibinfo{author}{\bibfnamefont{D.}~\bibnamefont{Martynov}},
  \bibinfo{author}{\bibfnamefont{R.}~\bibnamefont{Abbott}},
  \bibinfo{author}{\bibfnamefont{R.~X.} \bibnamefont{Adhikari}},
  \bibinfo{author}{\bibfnamefont{K.}~\bibnamefont{Arai}},
  \bibinfo{author}{\bibfnamefont{S.}~\bibnamefont{Ballmer}},
  \bibinfo{author}{\bibfnamefont{L.}~\bibnamefont{Barsotti}},
  \bibinfo{author}{\bibfnamefont{A.~F.} \bibnamefont{Brooks}},
  \bibinfo{author}{\bibfnamefont{R.~T.} \bibnamefont{DeRosa}},
  \bibinfo{author}{\bibfnamefont{S.}~\bibnamefont{Dwyer}},
  \bibnamefont{et~al.}, \bibinfo{journal}{Classical and Quantum Gravity}
  \textbf{\bibinfo{volume}{31}}, \bibinfo{pages}{245010}
  (\bibinfo{year}{2014}),
  \urlprefix\url{http://stacks.iop.org/0264-9381/31/i=24/a=245010}.

\bibitem[{\citenamefont{McCulloh}()}]{Louisiana_Active_Faults}
\bibinfo{author}{\bibfnamefont{R.~P.} \bibnamefont{McCulloh}},
  \emph{\bibinfo{title}{Active faults in east baton rogue parish, louisiana}},
  \bibinfo{note}{[Louisiana Geological Survey]},
  \urlprefix\url{https://www.lsu.edu/lgs/publications/products/Free_publications/EBR-faults.pdf}.

\bibitem[{\citenamefont{Veazey}(2018)}]{Rigzone}
\bibinfo{author}{\bibfnamefont{M.~V.} \bibnamefont{Veazey}},
  \bibinfo{journal}{Rigzone}  (\bibinfo{year}{2018}),
  \urlprefix\url{https://www.rigzone.com/news/new_player_enters_louisiana_austin_chalk-13-aug-2018-156605-article/}.

\bibitem[{\citenamefont{Richardson}(2018)}]{NaturalGasWorld}
\bibinfo{author}{\bibfnamefont{N.}~\bibnamefont{Richardson}},
  \bibinfo{journal}{Natural Gas World}  (\bibinfo{year}{2018}),
  \urlprefix\url{https://www.naturalgasworld.com/australis-gets-rig-for-us-tms-play-63952}.

\bibitem[{\citenamefont{of~Mineral~Resources}(2018)}]{DNR_Louisiana}
\bibinfo{author}{\bibfnamefont{L.~O.} \bibnamefont{of~Mineral~Resources}},
  \emph{\bibinfo{title}{Austin chalk play}} (\bibinfo{year}{2018}),
  \bibinfo{note}{[Online; accessed September 9, 2018)]},
  \urlprefix\url{http://www.dnr.louisiana.gov/assets/OMR/ACPforOMR.2018.pdf}.

\bibitem[{\citenamefont{Coughlin
  et~al.}(2017{\natexlab{b}})\citenamefont{Coughlin, Earle, Harms, Biscans,
  Buchanan, Coughlin, Donovan, Fee, Gabbard, Guy et~al.}}]{CoEa2017}
\bibinfo{author}{\bibfnamefont{M.}~\bibnamefont{Coughlin}},
  \bibinfo{author}{\bibfnamefont{P.}~\bibnamefont{Earle}},
  \bibinfo{author}{\bibfnamefont{J.}~\bibnamefont{Harms}},
  \bibinfo{author}{\bibfnamefont{S.}~\bibnamefont{Biscans}},
  \bibinfo{author}{\bibfnamefont{C.}~\bibnamefont{Buchanan}},
  \bibinfo{author}{\bibfnamefont{E.}~\bibnamefont{Coughlin}},
  \bibinfo{author}{\bibfnamefont{F.}~\bibnamefont{Donovan}},
  \bibinfo{author}{\bibfnamefont{J.}~\bibnamefont{Fee}},
  \bibinfo{author}{\bibfnamefont{H.}~\bibnamefont{Gabbard}},
  \bibinfo{author}{\bibfnamefont{M.}~\bibnamefont{Guy}}, \bibnamefont{et~al.},
  \bibinfo{journal}{Classical and Quantum Gravity}
  \textbf{\bibinfo{volume}{34}}, \bibinfo{pages}{044004}
  (\bibinfo{year}{2017}{\natexlab{b}}),
  \urlprefix\url{http://stacks.iop.org/0264-9381/34/i=4/a=044004}.

\bibitem[{\citenamefont{Resources}(2011)}]{Amelia_Resitivity}
\bibinfo{author}{\bibfnamefont{A.}~\bibnamefont{Resources}},
  \emph{\bibinfo{title}{Tuscaloosa marine shale: The significance of
  resistivity}}, \bibinfo{howpublished}{Online Presentations}
  (\bibinfo{year}{2011}),
  \urlprefix\url{http://ameliaresources.com/wp-content/uploads/2017/07/Amelia-Resources-LLC-WEBSITE-Tuscaloosa-Marine-Shale-RESISTIVITY-DISCUSSION-Mar-2011.pdf}.

\bibitem[{\citenamefont{Komatitsch and Vilotte}(1998{\natexlab{a}})}]{KoVi98}
\bibinfo{author}{\bibfnamefont{D.}~\bibnamefont{Komatitsch}} \bibnamefont{and}
  \bibinfo{author}{\bibfnamefont{J.~P.} \bibnamefont{Vilotte}},
  \bibinfo{journal}{bssa} \textbf{\bibinfo{volume}{88}}, \bibinfo{pages}{368}
  (\bibinfo{year}{1998}{\natexlab{a}}).

\bibitem[{\citenamefont{Tromp et~al.}(2008)\citenamefont{Tromp, Komatitsch, and
  Liu}}]{TrKoLi08}
\bibinfo{author}{\bibfnamefont{J.}~\bibnamefont{Tromp}},
  \bibinfo{author}{\bibfnamefont{D.}~\bibnamefont{Komatitsch}},
  \bibnamefont{and} \bibinfo{author}{\bibfnamefont{Q.}~\bibnamefont{Liu}},
  \bibinfo{journal}{Communications in Computational Physics}
  \textbf{\bibinfo{volume}{3}}, \bibinfo{pages}{1} (\bibinfo{year}{2008}).

\bibitem[{\citenamefont{Patera}(1984)}]{patera1984spectral}
\bibinfo{author}{\bibfnamefont{A.~T.} \bibnamefont{Patera}},
  \bibinfo{journal}{Journal of computational Physics}
  \textbf{\bibinfo{volume}{54}}, \bibinfo{pages}{468} (\bibinfo{year}{1984}).

\bibitem[{\citenamefont{Cockburn et~al.}(2011)\citenamefont{Cockburn,
  Karniadakis, and Shu}}]{Cockburn:2011:DGM:2408658}
\bibinfo{author}{\bibfnamefont{B.}~\bibnamefont{Cockburn}},
  \bibinfo{author}{\bibfnamefont{G.~E.} \bibnamefont{Karniadakis}},
  \bibnamefont{and} \bibinfo{author}{\bibfnamefont{C.-W.} \bibnamefont{Shu}},
  \emph{\bibinfo{title}{Discontinuous Galerkin Methods: Theory, Computation and
  Applications}} (\bibinfo{publisher}{Springer Publishing Company,
  Incorporated}, \bibinfo{year}{2011}), \bibinfo{edition}{1st} ed., ISBN
  \bibinfo{isbn}{3642640982, 9783642640988}.

\bibitem[{\citenamefont{Komatitsch and
  Vilotte}(1998{\natexlab{b}})}]{komatitsch1998spectral}
\bibinfo{author}{\bibfnamefont{D.}~\bibnamefont{Komatitsch}} \bibnamefont{and}
  \bibinfo{author}{\bibfnamefont{J.-P.} \bibnamefont{Vilotte}},
  \bibinfo{journal}{Bulletin of the seismological society of America}
  \textbf{\bibinfo{volume}{88}}, \bibinfo{pages}{368}
  (\bibinfo{year}{1998}{\natexlab{b}}).

\bibitem[{\citenamefont{Komatitsch and
  Tromp}(1999)}]{komatitsch1999introduction}
\bibinfo{author}{\bibfnamefont{D.}~\bibnamefont{Komatitsch}} \bibnamefont{and}
  \bibinfo{author}{\bibfnamefont{J.}~\bibnamefont{Tromp}},
  \bibinfo{journal}{Geophysical journal international}
  \textbf{\bibinfo{volume}{139}}, \bibinfo{pages}{806} (\bibinfo{year}{1999}).

\bibitem[{\citenamefont{Pellegrini and
  Roman}(1996)}]{pellegrini:1996:SSP:645560.658570}
\bibinfo{author}{\bibfnamefont{F.}~\bibnamefont{Pellegrini}} \bibnamefont{and}
  \bibinfo{author}{\bibfnamefont{J.}~\bibnamefont{Roman}}, in
  \emph{\bibinfo{booktitle}{Proceedings of the International Conference and
  Exhibition on High-Performance Computing and Networking}}
  (\bibinfo{publisher}{Springer-Verlag}, \bibinfo{address}{Berlin, Heidelberg},
  \bibinfo{year}{1996}), HPCN Europe 1996, pp. \bibinfo{pages}{493--498}, ISBN
  \bibinfo{isbn}{3-540-61142-8},
  \urlprefix\url{http://dl.acm.org/citation.cfm?id=645560.658570}.

\bibitem[{\citenamefont{USGS}(2017)}]{OkhCMT_USGS}
\bibinfo{author}{\bibnamefont{USGS}}, \emph{\bibinfo{title}{M 4.2 - 6km ENE of
  Edmond, Oklahoma}} (\bibinfo{year}{2017}),
  \urlprefix\url{https://earthquake.usgs.gov/earthquakes/eventpage/us2000a3y4#moment-tensor}.

\bibitem[{\citenamefont{Gaite et~al.}(2015)\citenamefont{Gaite, Villase{\~n}or,
  Iglesias, Herraiz, and Jim{\'e}nez-Munt}}]{gaite20153}
\bibinfo{author}{\bibfnamefont{B.}~\bibnamefont{Gaite}},
  \bibinfo{author}{\bibfnamefont{A.}~\bibnamefont{Villase{\~n}or}},
  \bibinfo{author}{\bibfnamefont{A.}~\bibnamefont{Iglesias}},
  \bibinfo{author}{\bibfnamefont{M.}~\bibnamefont{Herraiz}}, \bibnamefont{and}
  \bibinfo{author}{\bibfnamefont{I.}~\bibnamefont{Jim{\'e}nez-Munt}},
  \bibinfo{journal}{Solid Earth} \textbf{\bibinfo{volume}{6}},
  \bibinfo{pages}{271} (\bibinfo{year}{2015}).

\bibitem[{\citenamefont{Castagna and Backus}(1993)}]{castagna1993avo}
\bibinfo{author}{\bibfnamefont{J.~P.} \bibnamefont{Castagna}} \bibnamefont{and}
  \bibinfo{author}{\bibfnamefont{M.}~\bibnamefont{Backus}},
  \bibinfo{journal}{Offset-dependent reflectivity: Theory and practice of AVO
  analysis: SEG Investigations in Geophysics} \textbf{\bibinfo{volume}{8}},
  \bibinfo{pages}{3} (\bibinfo{year}{1993}).

\bibitem[{\citenamefont{Abernathy et~al.}(2011)\citenamefont{Abernathy,
  Acernese, Ajith, Allen, Amaro-Seoane, Andersson, Aoudia, Astone, Krishnan,
  Barack et~al.}}]{abernathy2011einstein}
\bibinfo{author}{\bibfnamefont{M.}~\bibnamefont{Abernathy}},
  \bibinfo{author}{\bibfnamefont{F.}~\bibnamefont{Acernese}},
  \bibinfo{author}{\bibfnamefont{P.}~\bibnamefont{Ajith}},
  \bibinfo{author}{\bibfnamefont{B.}~\bibnamefont{Allen}},
  \bibinfo{author}{\bibfnamefont{P.}~\bibnamefont{Amaro-Seoane}},
  \bibinfo{author}{\bibfnamefont{N.}~\bibnamefont{Andersson}},
  \bibinfo{author}{\bibfnamefont{S.}~\bibnamefont{Aoudia}},
  \bibinfo{author}{\bibfnamefont{P.}~\bibnamefont{Astone}},
  \bibinfo{author}{\bibfnamefont{B.}~\bibnamefont{Krishnan}},
  \bibinfo{author}{\bibfnamefont{L.}~\bibnamefont{Barack}},
  \bibnamefont{et~al.} (\bibinfo{year}{2011}).

\bibitem[{\citenamefont{Saulson}(1984)}]{saulson1984terrestrial}
\bibinfo{author}{\bibfnamefont{P.~R.} \bibnamefont{Saulson}},
  \bibinfo{journal}{Physical Review D} \textbf{\bibinfo{volume}{30}},
  \bibinfo{pages}{732} (\bibinfo{year}{1984}).

\bibitem[{\citenamefont{Hughes and Thorne}(1998)}]{hughes1998seismic}
\bibinfo{author}{\bibfnamefont{S.~A.} \bibnamefont{Hughes}} \bibnamefont{and}
  \bibinfo{author}{\bibfnamefont{K.~S.} \bibnamefont{Thorne}},
  \bibinfo{journal}{Physical Review D} \textbf{\bibinfo{volume}{58}},
  \bibinfo{pages}{122002} (\bibinfo{year}{1998}).

\bibitem[{\citenamefont{Harms}(2015)}]{Harms2015}
\bibinfo{author}{\bibfnamefont{J.}~\bibnamefont{Harms}},
  \bibinfo{journal}{Living Reviews in Relativity}
  \textbf{\bibinfo{volume}{18}}, \bibinfo{pages}{3} (\bibinfo{year}{2015}),
  ISSN \bibinfo{issn}{1433-8351},
  \urlprefix\url{https://doi.org/10.1007/lrr-2015-3}.

\bibitem[{\citenamefont{Coughlin et~al.}(2016)\citenamefont{Coughlin, Mukund,
  Harms, Driggers, Adhikari, and Mitra}}]{coughlin2016towards}
\bibinfo{author}{\bibfnamefont{M.}~\bibnamefont{Coughlin}},
  \bibinfo{author}{\bibfnamefont{N.}~\bibnamefont{Mukund}},
  \bibinfo{author}{\bibfnamefont{J.}~\bibnamefont{Harms}},
  \bibinfo{author}{\bibfnamefont{J.}~\bibnamefont{Driggers}},
  \bibinfo{author}{\bibfnamefont{R.}~\bibnamefont{Adhikari}}, \bibnamefont{and}
  \bibinfo{author}{\bibfnamefont{S.}~\bibnamefont{Mitra}},
  \bibinfo{journal}{Classical and Quantum Gravity}
  \textbf{\bibinfo{volume}{33}}, \bibinfo{pages}{244001}
  (\bibinfo{year}{2016}).

\bibitem[{\citenamefont{Coughlin et~al.}(2018)\citenamefont{Coughlin, Harms,
  Driggers, McManus, Mukund, Ross, Slagmolen, and
  Venkateswara}}]{coughlin2018implications}
\bibinfo{author}{\bibfnamefont{M.~W.} \bibnamefont{Coughlin}},
  \bibinfo{author}{\bibfnamefont{J.}~\bibnamefont{Harms}},
  \bibinfo{author}{\bibfnamefont{J.}~\bibnamefont{Driggers}},
  \bibinfo{author}{\bibfnamefont{D.~J.} \bibnamefont{McManus}},
  \bibinfo{author}{\bibfnamefont{N.}~\bibnamefont{Mukund}},
  \bibinfo{author}{\bibfnamefont{M.~P.} \bibnamefont{Ross}},
  \bibinfo{author}{\bibfnamefont{B.~J.~J.} \bibnamefont{Slagmolen}},
  \bibnamefont{and}
  \bibinfo{author}{\bibfnamefont{K.}~\bibnamefont{Venkateswara}},
  \bibinfo{journal}{Phys. Rev. Lett.} \textbf{\bibinfo{volume}{121}},
  \bibinfo{pages}{221104} (\bibinfo{year}{2018}),
  \urlprefix\url{https://link.aps.org/doi/10.1103/PhysRevLett.121.221104}.

\bibitem[{\citenamefont{Kohler et~al.}(2018)\citenamefont{Kohler, Cochran,
  Given, Guiwits, Neuhauser, Henson, Hartog, Bodin, Kress, Thompson
  et~al.}}]{kohler2018earthquake}
\bibinfo{author}{\bibfnamefont{M.~D.} \bibnamefont{Kohler}},
  \bibinfo{author}{\bibfnamefont{E.~S.} \bibnamefont{Cochran}},
  \bibinfo{author}{\bibfnamefont{D.}~\bibnamefont{Given}},
  \bibinfo{author}{\bibfnamefont{S.}~\bibnamefont{Guiwits}},
  \bibinfo{author}{\bibfnamefont{D.}~\bibnamefont{Neuhauser}},
  \bibinfo{author}{\bibfnamefont{I.}~\bibnamefont{Henson}},
  \bibinfo{author}{\bibfnamefont{R.}~\bibnamefont{Hartog}},
  \bibinfo{author}{\bibfnamefont{P.}~\bibnamefont{Bodin}},
  \bibinfo{author}{\bibfnamefont{V.}~\bibnamefont{Kress}},
  \bibinfo{author}{\bibfnamefont{S.}~\bibnamefont{Thompson}},
  \bibnamefont{et~al.}, \bibinfo{journal}{Seismological Research Letters}
  \textbf{\bibinfo{volume}{89}}, \bibinfo{pages}{99} (\bibinfo{year}{2018}).

\bibitem[{\citenamefont{Harms et~al.}(2015)\citenamefont{Harms, Ampuero,
  Barsuglia, Chassande-Mottin, Montagner, Somala, and
  Whiting}}]{harms2015transient}
\bibinfo{author}{\bibfnamefont{J.}~\bibnamefont{Harms}},
  \bibinfo{author}{\bibfnamefont{J.-P.} \bibnamefont{Ampuero}},
  \bibinfo{author}{\bibfnamefont{M.}~\bibnamefont{Barsuglia}},
  \bibinfo{author}{\bibfnamefont{E.}~\bibnamefont{Chassande-Mottin}},
  \bibinfo{author}{\bibfnamefont{J.-P.} \bibnamefont{Montagner}},
  \bibinfo{author}{\bibfnamefont{S.}~\bibnamefont{Somala}}, \bibnamefont{and}
  \bibinfo{author}{\bibfnamefont{B.}~\bibnamefont{Whiting}},
  \bibinfo{journal}{Geophysical Journal International}
  \textbf{\bibinfo{volume}{201}}, \bibinfo{pages}{1416} (\bibinfo{year}{2015}).

\bibitem[{\citenamefont{Montagner et~al.}(2016)\citenamefont{Montagner, Juhel,
  Barsuglia, Ampuero, Chassande-Mottin, Harms, Whiting, Bernard,
  Cl{\'e}v{\'e}d{\'e}, and Lognonn{\'e}}}]{montagner2016prompt}
\bibinfo{author}{\bibfnamefont{J.-P.} \bibnamefont{Montagner}},
  \bibinfo{author}{\bibfnamefont{K.}~\bibnamefont{Juhel}},
  \bibinfo{author}{\bibfnamefont{M.}~\bibnamefont{Barsuglia}},
  \bibinfo{author}{\bibfnamefont{J.~P.} \bibnamefont{Ampuero}},
  \bibinfo{author}{\bibfnamefont{E.}~\bibnamefont{Chassande-Mottin}},
  \bibinfo{author}{\bibfnamefont{J.}~\bibnamefont{Harms}},
  \bibinfo{author}{\bibfnamefont{B.}~\bibnamefont{Whiting}},
  \bibinfo{author}{\bibfnamefont{P.}~\bibnamefont{Bernard}},
  \bibinfo{author}{\bibfnamefont{E.}~\bibnamefont{Cl{\'e}v{\'e}d{\'e}}},
  \bibnamefont{and}
  \bibinfo{author}{\bibfnamefont{P.}~\bibnamefont{Lognonn{\'e}}},
  \bibinfo{journal}{Nature communications} \textbf{\bibinfo{volume}{7}},
  \bibinfo{pages}{13349} (\bibinfo{year}{2016}).

\bibitem[{\citenamefont{Commission}()}]{Arkansas_weblink}
\bibinfo{author}{\bibfnamefont{A.~O. .~G.} \bibnamefont{Commission}},
  \emph{\bibinfo{title}{Fayetteville shale information}},
  \bibinfo{note}{[Online; accessed September 9, 2018)]},
  \urlprefix\url{http://www.aogc.state.ar.us/sales/default.aspx}.

\bibitem[{\citenamefont{Evans}(1966)}]{osti_6805542}
\bibinfo{author}{\bibfnamefont{D.}~\bibnamefont{Evans}}, \bibinfo{journal}{Mt.
  Geol.; (United States)}  (\bibinfo{year}{1966}).

\bibitem[{\citenamefont{Hsieh and Bredehoeft}(1981)}]{hsieh1981reservoir}
\bibinfo{author}{\bibfnamefont{P.~A.} \bibnamefont{Hsieh}} \bibnamefont{and}
  \bibinfo{author}{\bibfnamefont{J.~D.} \bibnamefont{Bredehoeft}},
  \bibinfo{journal}{Journal of Geophysical Research: Solid Earth}
  \textbf{\bibinfo{volume}{86}}, \bibinfo{pages}{903} (\bibinfo{year}{1981}).

\bibitem[{\citenamefont{Commission}(2018)}]{Oklahoma_restrictions}
\bibinfo{author}{\bibfnamefont{O.~C.} \bibnamefont{Commission}},
  \emph{\bibinfo{title}{Earthquake response summary}} (\bibinfo{year}{2018}),
  \bibinfo{note}{[Online; accessed September 9, 2018)]},
  \urlprefix\url{http://www.occeweb.com/News/2018/05-30-18EARTHQUAKEACTIONSUMMARY.pdf}.

\bibitem[{\citenamefont{Hesthaven et~al.}(2007)\citenamefont{Hesthaven,
  Gottlieb, and Gottlieb}}]{hesthaven2007spectral}
\bibinfo{author}{\bibfnamefont{J.~S.} \bibnamefont{Hesthaven}},
  \bibinfo{author}{\bibfnamefont{S.}~\bibnamefont{Gottlieb}}, \bibnamefont{and}
  \bibinfo{author}{\bibfnamefont{D.}~\bibnamefont{Gottlieb}},
  \emph{\bibinfo{title}{Spectral methods for time-dependent problems}},
  vol.~\bibinfo{volume}{21} (\bibinfo{publisher}{Cambridge University Press},
  \bibinfo{year}{2007}).

\end{thebibliography}

\end{document}